\def\qed{\ifhmode\unskip\nobreak\fi\ifmmode\ifinner\else
 \hskip5pt\fi\fi\hbox{\hskip5pt\vrule width4pt
 height6pt depth1.5pt\hskip1pt}}

\documentclass[aip,amsmath,eqsecnum]{revtex4-1}

\draft 

\begin{document}

\title{High-energy analysis and Levinson's theorem for the selfadjoint matrix Schr\"odinger operator on the half line} 

\author{Tuncay Aktosun}
\email[]{aktosun@uta.edu}
\affiliation{Department of Mathematics, University of Texas at Arlington, Arlington, TX 76019-0408, USA}
\author{Ricardo Weder}
\email[]{weder@unam.mx}
\thanks{Fellow Sistema Nacional de Investigadores}
\affiliation{Departamento de F\'\i sica Matem\'atica, Instituto de Investigaciones en
Matem\'aticas Aplicadas y en Sistemas, Universidad Nacional Aut\'onoma de M\'exico, Apartado Postal 20-726, IIMAS-UNAM, M\'exico DF 01000, M\'exico}

\date{\today}

\begin{abstract}
The matrix Schr\"odinger equation
with a selfadjoint matrix potential is considered on the half
line with the general selfadjoint boundary condition at
the origin. When the matrix potential is integrable,
the high-energy asymptotics are established for the related
Jost matrix, the inverse of the Jost matrix, and the scattering matrix.
Under the additional assumption that the matrix potential
has a first moment, Levinson's theorem
is derived, relating the number of bound states
to the change in the argument of the determinant of
the scattering matrix.
\end{abstract}

\pacs{2.30.Zz 3.65.-w 3.65.Ge 3.65.Nk}

\maketitle 

\section{Introduction}
Consider the matrix Schr\"odinger equation on the half line
\begin{equation}
-\psi''+V(x)\,\psi=k^2\psi,\qquad x\in{\bf R}^+,\label{1.1}
\end{equation}
where ${\bf R}^+:=(0,+\infty),$ the prime denotes the derivative with respect to
the spatial coordinate $x,$ and
the potential $V$ is a $n\times n$ selfadjoint matrix-valued function integrable in $x.$
The integrability $V\in L^1({\bf R}^+)$ means that
each entry of the matrix $V$ is Lebesgue measurable on
${\bf R}^+$ and
\begin{equation}\int_0^\infty
dx\,||V(x)||<+\infty,\label{1.2}\end{equation} where $||V(x)||$ denotes a
matrix norm. Since all matrix norms are
equivalent, without loss of generality, we can use the matrix norm defined as
$$||V(x)||:=\max_{l}\sum_{s=1}^n |V_{ls}(x)|,\qquad l=1,\dots,n,$$
where $V_{ls}(x)$ denotes the $(l,s)$-entry of the matrix $V(x).$
Clearly, a matrix-valued function is integrable in $x$
if and only if each entry of that matrix belongs
to $L^1({\bf R}^+).$

Note that $V$ is not assumed to be real
valued but is assumed to be selfadjoint,
i.e.
\begin{equation}V(x)=V(x)^\dagger,\qquad x\in{\bf R}^+,\label{1.3}\end{equation}
where the dagger denotes the matrix adjoint (complex
conjugate and matrix transpose).
The wavefunction $\psi(k,x)$ appearing in (1.1)
will be either an $n\times n$ matrix-valued function
or it will be a column vector with $n$
components.

When the analysis of (1.1) is considered at or near $k=0,$
in addition to the integrability we require that
the potential $V$ has a first moment, i.e.
\begin{equation}\int_0^\infty
dx\,x\,||V(x)||<+\infty.\label{1.4}\end{equation}
All the results in our paper are valid
under the assumption that the potential $V$
is selfadjoint and belongs to $L^1_1({\bf R}^+),$ i.e.
\begin{equation}\int_0^\infty
dx\,(1+x)\,||V(x)||<+\infty.\label{1.5}\end{equation}

We are interested in studying (1.1) with a selfadjoint potential $V$
in $L^1_1({\bf R}^+)$ under the general
selfadjoint boundary condition at $x=0.$
As indicated in Ref.~\citenum{2}, without loss of generality,
it is convenient to state the general
selfadjoint boundary condition at $x=0$ for (1.1) in terms of
constant $n\times n$ matrices $A$ and $B$ such that
\begin{equation}-B^\dagger\psi(0)+A^\dagger\psi'(0)=0,\label{1.6}\end{equation}
\begin{equation}-B^\dagger A+A^\dagger B=0,\label{1.7}\end{equation}
\begin{equation}A^\dagger A+B^\dagger B>0,\label{1.8}\end{equation}
i.e. $A^\dagger B$ is selfadjoint and
the selfadjoint matrix $(A^\dagger
A+B^\dagger B)$ is positive.
There are various equivalent formulations \cite{2,12,13,14,15,16} of
the general
selfadjoint boundary condition at $x=0.$
Let us also mention that it is possible to use some transformations on
$A$ and $B$ without affecting (1.7) and (1.8). In Section~4 we will elaborate on two
such transformations,
namely $(A,B)\mapsto (AT,BT),$ a right multiplication by
an invertible matrix $T,$ and a unitary transformation
$(A,B)\mapsto (M^\dagger AM,M^\dagger BM),$ where $M$ is
a unitary matrix. A combination of such transformations as given in (4.10)
will turn out to be useful.

We have two primary goals in this paper
for the Schr\"odinger equation
(1.1) with the selfadjoint
boundary condition (1.6)-(1.8) when the potential $V$ satisfies (1.3) and (1.5).
Our first primary goal is, even when $V$ satisfies the weaker condition (1.2) instead of
(1.5), to establish
the large-$k$
asymptotics of various quantities related to (1.1) such as some relevant
scattering solutions, the Jost matrix, the inverse of the Jost
matrix, and the scattering matrix. Our second primary goal is,
under the additional assumption (1.4), to derive so-called Levinson's
theorem, namely to obtain the relationship between the number of
bound states and the change in the phase of the determinant of the
scattering matrix.

A bound state
corresponds to a square-integrable
column-vector solution to (1.1) satisfying the boundary condition (1.6)-(1.8).
The selfadjointness (1.3) of $V$ and the selfadjoint boundary condition (1.6)-(1.8)
assure that the corresponding Schr\"odinger operator
is selfadjoint on $L^2({\bf R}^+),$ and hence its eigenvalues must be real.
When $k^2\ge 0,$ it turns out that there are no square-integrable column-vector solutions to (1.1).
As a result, a bound state, if it exists, occurs only when $k^2$ is negative,
or equivalently when $k$ appearing in
(1.1) is on the positive imaginary axis in the complex plane ${\bf C}.$
Thanks to the restriction (1.5), the number of such $k$-values turns out to be finite.
We will see that, at each
$k$-value corresponding to a bound state,
the number of linearly independent square-integrable column-vector
solutions (i.e. the multiplicity of the corresponding bound state) cannot exceed $n.$
The number of bound states is defined as
the number of bound states including the multiplicities.

The large-$k$ analysis and Levinson's theorem for
(1.1) with the selfadjoint boundary condition (1.6)-(1.8) are
relevant in the study of the corresponding
direct and inverse scattering problems. The direct scattering
problem for (1.1) is to determine the scattering matrix and the
bound-state information when the matrix potential $V$ and the
selfadjoint boundary condition are known. On the other hand,
the inverse scattering problem is to recover the potential and
the boundary condition from an appropriate set of scattering
data.

Our paper
complements the study \cite{1} by Agranovich and Marchenko, where
the large-$k$ asymptotics and Levinson's theorem are provided only under the
Dirichlet boundary condition. Our study also complements
the study \cite{12,13,14} by Harmer (see also Ref.~\citenum{5}), where the general selfadjoint
boundary condition is used
but the large-$k$ asymptotics of the scattering matrix is obtained
by only providing the leading term with the remaining terms
as specified as $o(1)$ as $k$ becomes large. In our paper,
we not only provide the leading term but we also specify the
next order term and establish the
large-$k$ asymptotics up to $O(1/k^2),$ which is crucial in establishing the
Fourier transforms of various quantities relevant
to the corresponding inverse scattering problem.

Our current paper also complements our own recent study,\cite{2} where
the rigorous small-$k$ analysis for (1.1) is provided
with the general selfadjoint boundary condition (1.6)-(1.8).
The small-$k$ analysis in Ref.~\citenum{2} is crucial in the derivation of
Levinson's theorem in our current paper. In fact, the only reason for needing
(1.5) rather than merely (1.2)
in our current paper is because of the fact that the small-$k$ asymptotics
are also needed to establish Levinson's theorem, and those asymptotics require (1.5).

The half-line matrix Schr\"odinger equation (1.1)
has applications in quantum mechanical scattering involving
particles of internal structures such as spins,
in scattering on
graphs, \cite{4,6,7,8,9,10,11,17,18,19,20,21} and in quantum wires. \cite{15,16}
The consideration of
the general selfadjoint boundary condition at $x=0$ given in (1.6)-(1.8)
rather than the Dirichlet boundary condition $\psi(0)=0$ is relevant.
For example, the half-line matrix
Schr\"odinger equation (1.1) describes the behavior of $n$ connected very thin
quantum wires forming a one-vertex graph with open ends, and
the boundary condition (1.6)-(1.8)
imposes certain restrictions at the vertex.
Such a problem is useful in designing elementary gates
in quantum computing and nanotubes for microscopic electronic devices,
where, for example, strings of atoms may form a star-shaped graph.
For details we refer the reader
to Refs.~\citenum{15} and \citenum{16} and the references therein.

Our paper is organized as follows.
In Section~2 we introduce various $n\times n$ matrix
solutions to (1.1) that are needed later on.
In Section~3 we introduce the Jost matrix $J(k)$ and the scattering matrix $S(k)$ and state
some of their properties relevant to our study.
In Section~4 we introduce two key transformations
on the boundary condition (1.6)-(1.8) and analyze
how those transformations affect the Jost matrix and the scattering matrix.
In Section~5 we provide the relevant properties of
$J_0(k)$ and $S_0(k),$ the Jost and scattering matrices
corresponding to $V\equiv 0;$ those properties
are crucial in understanding similar properties when the potential in nonzero.
In Section~6 we analyze the behavior of the Jost matrix at $k=0,$ which is needed
in establishing Levinson's theorem.
In Section~7 we analyze the large-$k$ asymptotics of the Jost and scattering matrices.
In Section~8 we analyze the bound states and the properties of the Jost matrix
related to bound states.
Finally, in Section~9 we establish Levinson's theorem, showing
how the change in the phase of the determinant of the scattering matrix is related
to the number of bound states.

\section{Scattering solutions}

In this section we introduce certain $n\times n$ matrix
solutions to (1.1) and
recall some of their properties relevant to our study.
We use ${\bf C}^+$ to denote
the upper-half complex plane and
${\bf R}$ for the real axis, and we let ${\overline{{\bf C}^+}}:={\bf C}^+\cup{\bf R}.$
Recall that we assume that the potential $V$ appearing in
(1.1) satisfies (1.3) and (1.5).

The Jost solution to
(1.1) is the $n\times n$ matrix solution
satisfying, for $k\in{\overline{{\bf C}^+}}\setminus\{0\},$ the asymptotics
\begin{equation}f(k,x)=e^{ikx}[I_n+o(1/x)],\quad
f'(k,x)=ik\,e^{ikx}[I_n+o(1/x)],\qquad x\to+\infty,\label{2.1}\end{equation}
where $I_n$ denotes the $n\times n$ identity matrix. It
satisfies the integral equation
\begin{equation}f(k,x)=e^{ikx}I_n+\displaystyle\frac{1}{k}\int_x^\infty dy\,\sin k(y-x)\,V(y)\,f(k,y),\label{ 2.2}\end{equation}
and it is known \cite{1,2} that
$f(k,x)$ and $f'(k,x)$ are analytic
in $k\in{\bf C}^+$
and continuous in $k\in{\overline{{\bf C}^+}}$ for each fixed $x.$
We remark that $f(k,x)$ corresponds to the quantity $E(-k,x)$ described
on p. 28 of Ref.~\citenum{1}. From (2.1) it is seen that each of the $n$ columns of
$f(k,x)$ exponentially decays to zero as $x\to+\infty$ for each fixed
$k\in{\bf C}^+.$

The matrix Schr\"odinger equation (1.1) has the $n\times n$ matrix solution $g(k,x)$
satisfying, for each $k\in{\overline{{\bf C}^+}}\setminus\{0\},$ the asymptotics
\begin{equation}g(k,x)=e^{-ikx}[I_n+o(1/x)],\quad
g'(k,x)=-ik\,e^{-ikx}[I_n+o(1/x)],\qquad x\to+\infty,\label{2.3}\end{equation}
and $g(k,x)$ corresponds to the quantity $E^{(1)}(-k,x)$ described
on p. 28 of Ref.~\citenum{1}. It is known \cite{1} that
$g(k,x)$ and $g'(k,x)$ are analytic
in $k\in{\bf C}^+$
and continuous in $k\in{\overline{{\bf C}^+}}\setminus\{0\}$ for each fixed $x.$ From
(2.3) it is seen that each of the $n$ columns of
$g(k,x)$ exponentially grows as $x\to+\infty$ for each fixed
$k\in{\bf C}^+.$

As indicated on p. 28 of Ref.~\citenum{1}, for each
$k\in{\overline{{\bf C}^+}}\setminus\{0\},$ the combined $2n$ columns of $f(k,x)$ and $g(k,x)$
form a fundamental set of solutions to (1.1), and hence
any column-vector solution $\omega(k,x)$ to (1.1) can be written as
\begin{equation}\omega(k,x)=f(k,x)\,\xi+g(k,x)\,\eta,\label{2.4}\end{equation}
for some constant column vectors $\xi$ and $\eta$ in ${\bf C}^n.$

The regular solution $\varphi(k,x)$ is the $n\times n$ matrix solution to (1.1)
satisfying the initial conditions
\begin{equation}\varphi(k,0)=A,\quad \varphi'(k,0)=B,\label{2.5}\end{equation}
where $A$ and $B$ are the matrices appearing in (1.6)-(1.8). For each fixed
$x\in{\bf R}^+,$ it is known that $\varphi(k,x)$ is entire in $k$
in the complex plane ${\bf C}.$
Note that
\begin{equation}\varphi(-k,x)=\varphi(k,x),\qquad k\in{\bf C},\quad x\in{\bf R}^+,\label{2.6}\end{equation}
because $k$ appears as $k^2$ in (1.1) and
the initial values given in (2.5)
are independent of $k.$

We will use
$[F;G]:=FG'-F'G$ for the Wronskian and use an asterisk to
denote complex conjugation. It can be verified directly
that for any two $n\times n$ solutions $\phi(k,x)$ and
$\psi(k,x)$ to (1.1), each of
the Wronskians
$[\phi(k^*,x)^\dagger;\psi(k,x)]$ and
$[\phi(-k^*,x)^\dagger;\psi(k,x)]$ is independent of $x.$
By evaluating the values of the Wronskians at two different
$x$-values, say $x=0$ and $x=+\infty,$
we can obtain various useful identities.
For example, we have
\begin{equation}[f(\pm k,x)^\dagger;f(\pm k,x)]=
\pm 2ikI_n,\qquad k\in{\bf R},\label{2.7}\end{equation}
\begin{equation}[f(-k^*,x)^\dagger;f(k,x)]=
0,\qquad k\in{\overline{{\bf C}^+}}.\label{2.8}\end{equation}
Let us add that, for each fixed $x\in{\bf R}^+,$ if a solution
$\phi(k,x)$ is analytic in $k\in{\bf C}^+$ then
$\phi(-k^*,x)^\dagger$ becomes an analytic function of $k$ in ${\bf C}^+;$
on the other hand, $\phi(k^*,x)^\dagger$ becomes an analytic function of
$k$ in the lower-half complex plane ${\bf C}^-.$

\section{The Jost matrix and the scattering matrix}

In this section we introduce the Jost matrix $J(k)$ and
the scattering matrix $S(k)$ for (1.1) with a selfadjoint matrix potential $V$ in
$L^1_1({\bf R}^+)$ and with the selfadjoint boundary condition
(1.6)-(1.8).
We also recall or establish
some of their properties relevant
to our study.

The Jost matrix $J(k)$ is defined in terms of a
Wronskian as
\begin{equation}J(k):=[f(-k^*,x)^\dagger;\varphi(k,x)],\qquad k\in{\overline{{\bf C}^+}},\label{3.1}\end{equation}
where
$f(k,x)$ is the Jost solution appearing in (2.1) and
$\varphi(k,x)$ is the regular solution appearing in (2.5).
Since the Wronskian in (3.1) is independent of $x,$ by evaluating
its value at $x=0,$ with the help of (2.5) we get
\begin{equation}J(k)=f(-k^*,0)^\dagger B-f'(-k^*,0)^\dagger A,\label{3.2}\end{equation}
where $A$ and $B$ are the matrices appearing in (1.6)-(1.8).
Note that the domain of $J(k)$ is ${\overline{{\bf C}^+}}$ because $f(-k^*,0)^\dagger$
and $f'(-k^*,0)^\dagger$ are analytic in $k\in{{\bf C}^+}$ and
continuous in $k\in{\overline{{\bf C}^+}},$ as $f(k,0)$ and $f'(k,0)$
are analytic in $k\in{{\bf C}^+}$ and
continuous in $k\in{\overline{{\bf C}^+}}.$

We quote the following fundamental results from Ref.~\citenum{2} regarding the Jost matrix and its inverse
and their
small-$k$ behavior. We refer the reader to Theorem~4.1 and Theorem~6.3 of Ref.~\citenum{2}
for further details.
These results are later needed
in the derivation of Levinson's theorem.

{\bf Theorem 3.1:} {\it Consider the selfadjoint matrix
Schr\"odinger operator with the selfadjoint boundary condition
(1.6)-(1.8) and with the potential $V$ satisfying (1.3) and (1.5).
Then:}

\begin{itemize}
\item[(a)] {\it $J(k)$ is analytic in
${{\bf C}^+}$ and continuous in ${\overline{{\bf C}^+}}.$}

\item[(b)] {\it As $k\to 0$ in ${\overline{{\bf C}^+}}$ we have}
\begin{equation}J(k)={\mathcal{S}} P_2^{-1}
\left[ \begin{array} {cc}k {\mathcal{A}}_1+o(k)&k {\mathcal{B}}_1+o(k)\\
k {\mathcal{C}}_1+o(k)&{\mathcal{D}}_0+o(1)\end{array}\right] P_1 {\mathcal{S}}^{-1},\label{3.3}\end{equation}
\item[]{\it where
$P_1$ and $P_2$ are some permutation matrices,
 and ${\mathcal{S}}$ is an invertible matrix,
and
${\mathcal{A}}_1,$ ${\mathcal{B}}_1,$ ${\mathcal{C}}_1,$ and ${\mathcal{D}}_0$
are some constant matrices of sizes $\mu\times \mu,$ $\mu\times(n-\mu),$
$(n-\mu)\times\mu,$ and $(n-\mu)\times(n-\mu),$ respectively, in such a way
that ${\mathcal{A}}_1$ and ${\mathcal{D}}_0$ are both invertible. Here,
$\mu$ is the geometric multiplicity
of the zero eigenvalue of the zero-energy
 Jost matrix $J(0).$}

\item[(c)] {\it $J(k)$ is invertible for $k\in{\bf R}\setminus\{0\},$ and in fact
$J(k)^{-1}$ is continuous for $k\in{\bf R}\setminus\{0\}.$}

\item[(d)] {\it As $k\to 0$ in ${\overline{{\bf C}^+}},$ we have}
\begin{equation}J(k)^{-1}={\mathcal{S}} P_1
\left[ \begin{array} {cc}\displaystyle\frac{1}{k}\, {\mathcal{A}}_1^{-1}[I_\mu+o(1)]&
- {\mathcal{A}}_1^{-1}{\mathcal{B}}_1  {\mathcal{D}}_0^{-1}+o(1)\\
-{\mathcal{D}}_0^{-1}{\mathcal{C}}_1{\mathcal{A}}_1^{-1}+o(1)&
{\mathcal{D}}_0^{-1}+o(1)\end{array}\right] P_2{\mathcal{S}}^{-1}.\label{3.4}\end{equation}
{\it Hence, $J(k)^{-1}$ is either continuous at $k=0,$ or it has a simple pole at $k=0.$
In particular, $kJ(k)^{-1}$
has a well-defined limit at $k=0.$}

\end{itemize}

The following result is already known, but we provide a brief proof for the reader's benefit,
as the information contained in the proof
is relevant to our study.

{\bf Proposition 3.2;} {\it Consider the selfadjoint matrix
Schr\"odinger operator with the selfadjoint boundary condition
(1.6)-(1.8) and with the potential $V$ satisfying (1.3) and (1.5).
Then, the regular solution
$\varphi(k,x)$ can be expressed in terms of the Jost solution
$f(k,x)$ and the Jost matrix $J(k)$ as}
\begin{equation}\varphi(k,x)={\displaystyle}\frac{1}{2ik}\,
f(k,x)\,J(-k)-{\displaystyle}\frac{1}{2ik}\,f(-k,x)\,J(k),\qquad k\in{\bf R}\setminus\{0\}.\label{ 3.5}\end{equation}

{\it{Proof:}} With the help of (2.1) we see that the
combined $2n$ columns of $f(k,x)$ and $f(-k,x)$
form a fundamental set of column-vector
solutions to (1.1) for any $k\in{\bf R}\setminus\{0\},$ and hence
we can write the regular solution $\varphi(k,x)$ as
\begin{equation}\varphi(k,x)=f(k,x)\,C_1(k)+f(-k,x)\,C_2(k),\qquad k\in{\bf R}\setminus\{0\},\label{ 3.6}\end{equation}
for some $n\times n$ matrices $C_1(k)$ and $C_2(k)$ depending only on $k$ but not on $x.$
Using (3.6) in the
Wronskians $[f(\pm k,x)^\dagger;\varphi(k,x)],$ with the help of
(2.7), (2.8), and (3.1) we obtain
$$C_1(k)={\displaystyle}\frac{1}{2ik}\,J(-k),\quad C_2(k)=-{\displaystyle}\frac{1}{2ik}\,J(k),$$
yielding (3.5) for real nonzero values of $k.$ \qed

Let us define the $n\times n$ physical solution $\Psi(k,x)$ to (1.1)
as
\begin{equation}\Psi(k,x):=-2ik\,\varphi(k,x)\,J(k)^{-1},\label{3.7}\end{equation}
where $\varphi(k,x)$ is the regular solution appearing in
(2.5) and $J(k)$ is the Jost matrix defined in (3.1).
The scattering matrix $S(k)$ is defined as \cite{2,12,13,14}
\begin{equation}S(k):=-J(-k)\,J(k)^{-1},\qquad k\in{\bf R}.\label{3.8}\end{equation}
As we elaborate in Section~8, $J(k)$ can uniquely be defined
only up to a multiplication
on the right by an invertible
constant matrix.
On the other
hand, as seen from (3.8), such a postmultiplication of
$J(k)$ by an invertible matrix does not affect $S(k).$ Hence, $S(k)$
is uniquely determined by the potential $V$ and the boundary condition (1.6),
independently of the particular parametrization used in (1.6)-(1.8).
In general,
$S(k)$ is defined only for real $k$ because $J(-k)$
in general cannot be extended from $k\in{\bf R}$ to $k\in{{\bf C}^+}.$
The continuity of $S(k)$ at $k=0$ has recently been established. \cite{2}
Even when $J(k)^{-1}$ may not exist at $k=0,$
it has been shown \cite{2} that
the product on the right-hand side in (3.8) has a well-defined
limit as $k\to 0$ in ${\bf R},$ and hence the domain of $S(k)$ is $k\in{\bf R}.$
The small-$k$ behavior
of $S(k)$ is quoted from Ref.~\citenum{2} in the following and hence a proof is omitted.

{\bf Proposition 3.3:} {\it Consider the selfadjoint matrix
Schr\"odinger operator with the selfadjoint boundary condition
(1.6)-(1.8) and with the potential $V$ satisfying (1.3) and (1.5). Then,
the scattering matrix
$S(k)$ defined in (3.8) is continuous for $k\in{\bf R}$ including $k=0,$ and
we have $S(k)=S(0)+o(1)$ as $k\to 0$ in ${\bf R}$ with}
\begin{equation}S(0)={\mathcal{S}} P_2^{-1}
\left[ \begin{array} {cc} I_\mu& 0\\
2{\mathcal{C}}_1{\mathcal{A}}_1^{-1}&-I_{n-\mu}\end{array}\right]
P_2  {\mathcal{S}}^{-1},\label{3.9}\end{equation}
{\it where
$\mu$ is the geometric multiplicity
of the zero eigenvalue of the zero-energy
 Jost matrix $J(0),$
$P_2$ is an $n\times n$ permutation matrix,
${\mathcal{S}}$ is an $n\times n$ constant invertible matrix,
${\mathcal{A}}_1$ is a $\mu\times \mu$ constant invertible matrix,
and ${\mathcal{C}}_1$ is an $(n-\mu)\times\mu$ constant matrix.}

We note that the quantities $\mu,$ $P_2,$ ${\mathcal{S}},$
${\mathcal{A}}_1,$ and ${\mathcal{C}}_1$ appearing in (3.9) are the same
as those appearing in (3.3) and (3.4).

{\bf Proposition 3.4:} {\it Consider the selfadjoint matrix
Schr\"odinger operator with the selfadjoint boundary condition
(1.6)-(1.8) and with the potential $V$ satisfying (1.3) and (1.5). Then for each $x\in{\bf R}^+$ the physical solution given in (3.7) is continuous for $k\in{\bf R}$ and can be written as}
\begin{equation}\Psi(k,x):=f(-k,x)+f(k,x)\,S(k),\qquad k\in{\bf R},\label{3.10}\end{equation}
{\it where $S(k)$ is the scattering matrix defined in (3.8).}

{\it{Proof:}} Using (3.5) and (3.8) in (3.7) we get (3.10) for $k\in{\bf R}\setminus\{0\}.$
 From the continuity \cite{2} of $f(k,x)$ and $S(k)$ for $k\in{\bf R}$ including $k=0,$ it follows that
(3.10) also holds at $k=0,$ and hence $\Psi(k,x)$ is continuous in $k\in{\bf R}$ for each
fixed $x\in{\bf R}^+.$ We can verify the continuity of
$\Psi(k,x)$ in $k\in{\bf R}$ in an alternate way.
As stated in Theorem~3.1(d),
even though $J(k)^{-1}$ may not exist at $k=0,$ the
quantity $kJ(k)^{-1}$ is continuous for $k\in{\bf R}.$
We recall that $\varphi(k,x)$ is entire in $k$ for
each $x\in{\bf R}^+.$
Thus, for each fixed $x\in{\bf R}^+,$
the physical solution $\Psi(k,x)$ defined in (3.7)
is continuous in $k\in{\bf R}.$ \qed

Some useful properties of the scattering matrix $S(k)$ are provided in the following
proposition. Thanks to the recent result \cite{2} on the small-$k$ limit
of $S(k),$ the properties listed below
hold for any real $k,$ including $k=0.$

{\bf Proposition 3.5:} {\it Consider the selfadjoint matrix
Schr\"odinger operator with the selfadjoint boundary condition
(1.6)-(1.8) and with the potential $V$ satisfying (1.3) and (1.5).
Then the scattering matrix $S(k)$ defined
in (3.8) is unitary for $k\in{\bf R}$ and satisfies}
\begin{equation}S(-k)=S(k)^{-1}=S(k)^\dagger,\qquad k\in{\bf R}.\label{3.11}\end{equation}

{\it{Proof:}}
Using (3.10) in the Wronskian $[\Psi(k,x)^\dagger;\Psi(k,x)],$
with the help of (2.7) and (2.8) we get
\begin{equation}[\Psi(k,x)^\dagger;\Psi(k,x)]=-2ikI_n+2ik\,S(k)^\dagger S(k),\qquad k\in{\bf R}.
\label{3.12}\end{equation}
On the other hand, using (3.7) in the same Wronskian, with the help of
(1.7) and (2.5) we obtain
\begin{equation}[\Psi(k,x)^\dagger;\Psi(k,x)]=-(2ik)^2 [J(k)^\dagger]^{-1}
\left(A^\dagger B-B^\dagger A\right)J(k)^{-1}=0,\qquad k\in{\bf R}\setminus\{0\}.
\label{3.13}\end{equation}
In fact, (3.13) holds also at $k=0$ by letting $k\to 0$ and noting that
$kJ(k)^{-1}$ has a well-defined limit \cite{2}
as $k\to 0$ in ${\overline{{\bf C}^+}},$ as stated in Theorem~3.1(d).
Comparing (3.12) and (3.13) we then get
$S(k)^\dagger S(k)=I_n$ for $k\in{\bf R},$ which
yields $S(k)^{-1}=S(k)^\dagger$ for $k\in{\bf R}.$
To establish $S(-k)=S(k)^\dagger$ for $k\in{\bf R}$ we proceed by evaluating
the Wronskian $[\Psi(-k,x)^\dagger;\Psi(k,x)]$ in two different
ways. First, using (3.10) in that Wronskian, with the help of
(2.7) and (2.8) we get
\begin{equation}[\Psi(-k,x)^\dagger;\Psi(k,x)]=2ik\left(S(k)-S(-k)^\dagger\right),
\qquad k\in{\bf R}.
\label{3.14}\end{equation}
On the other hand, using (3.7) in the same Wronskian, with the
help of
(1.7), (2.5), and (2.6) we obtain
\begin{equation}[\Psi(-k,x)^\dagger;\Psi(k,x)]=(2ik)^2 [J(-k)^\dagger]^{-1}
\left(A^\dagger B-B^\dagger A\right)J(k)^{-1}=0,\qquad k\in{\bf R}\setminus\{0\}.\label{ 3.15}\end{equation}
For the same reason (3.13) holds at $k=0,$ we conclude that
(3.15) also holds at $k=0.$ Thus, comparing (3.14) and (3.15)
we conclude that $S(-k)=S(k)^\dagger$ for $k\in{\bf R}.$ \qed

\section{Transformations}

In this section we remark how the Jost solution and the
regular solution to (1.1), the Jost matrix, and the scattering
matrix change if the matrices
$A$ and $B$ used in the parametrization of the boundary condition
(1.6)-(1.8) undergo a transformation
without affecting (1.7) and (1.8). In particular, we consider
a multiplication on the right by an invertible matrix,
a unitary transformation by a unitary matrix, and a combination of those two
transformations. The results will be useful in
analyzing the large-$k$ asymptotics of various
quantities and in the derivation of Levinson's theorem.

{\bf Proposition 4.1:} {\it Consider the selfadjoint matrix
Schr\"odinger operator with the selfadjoint boundary condition
(1.6)-(1.8) and with the potential $V$ satisfying (1.3) and (1.5).
Let $A$ and $B$ be the matrices appearing in
(1.6)-(1.8), $f(k,x)$ be the Jost solution to (1.1)
satisfying (2.1), $\varphi(k,x)$ be the regular solution to (1.1)
satisfying (2.5), $J(k)$ be the Jost matrix defined in (3.1), and
$S(k)$ be the scattering matrix defined in (3.8). Then:}

\begin{itemize}
\item[(a)] {\it Under the transformation $V\mapsto V$ and
$(A,B)\mapsto (AT,BT),$ where $T$ is an invertible
$n\times n$ matrix, we have}
$$(f,\varphi,J,S)\mapsto (f,\varphi T, JT,S).$$

\item[(b)] {\it Under the unitary transformation $V\mapsto M^\dagger V M$ and
$(A,B)\mapsto (M^\dagger A M,M^\dagger B M),$ where $M$ is a unitary
$n\times n$ matrix, we have}
$$(f,\varphi,J,S)\mapsto (M^\dagger f M,M^\dagger \varphi M,M^\dagger J M,M^\dagger S M).$$

\item[(c)] {\it Under the unitary transformation $V\mapsto M^\dagger V M$
with a unitary matrix $M$
and
the combination of three consecutive
transformation $(A,B)\mapsto (M^\dagger AT_1 MT_2,M^\dagger B T_1 MT_2),$
first by a right multiplication by an invertible matrix
$T_1,$ then by the unitary transformation with $M,$
followed by a right multiplication by an invertible matrix
$T_2,$ we have}
\begin{equation}(f,\varphi,J,S)\mapsto (M^\dagger f M,M^\dagger \varphi
T_1MT_2,M^\dagger JT_1 MT_2,M^\dagger S M).\label{4.1}\end{equation}
\end{itemize}

{\it{Proof:}} The proof is obtained by direct verification
and by checking that the boundary condition (1.6)-(1.8),
the Schr\"odinger equation (1.1), and the relevant conditions and definitions in
(2.1), (2.5), (3.1), and (3.8) all remain satisfied. Finally, the transformation in (c) is obtained
 from the results in (a) and (b). \qed

We note that the transformation $V\mapsto V$ and $(A,B)\mapsto (AT,BT)$ with
an invertible matrix $T$ is just a change of
parametrization in the boundary condition (1.6)-(1.8).
On the other hand, the unitary
transformation $V\mapsto M^\dagger V M$
and $(A,B)\mapsto (M^\dagger A M,M^\dagger B M)$
with a unitary matrix $M$
is a change of representation in the sense of
quantum mechanics.

Motivated by the general selfadjoint boundary condition \cite{3,22,23} in the scalar case, i.e.
the case with $n=1,$
we are interested in going from the pair $A$ and $B$ appearing
in the selfadjoint boundary condition (1.6)-(1.8) to
the special pair $\tilde A$ and $\tilde B,$ where
we have defined
\begin{equation}\tilde A:=-\text{diag}\{\sin\theta_1,\dots,\sin\theta_n\},
\quad \tilde B:=\text{diag}\{\cos\theta_1,\dots,\cos\theta_n\},\label{4.2}\end{equation}
with the real parameters $\theta_j$ taking
values in the interval $(0,\pi].$
The special case
$\theta_j=\pi$ corresponds to the Dirichlet boundary condition and the case
$\theta_j=\pi/2$ corresponds to the Neumann boundary condition.
We assume that there
are $n_{\text N}$ values with $\theta_j=\pi/2$ and $n_{\text D}$ values with $\theta_j=\pi,$
and hence there are $n_{\text M}$ remaining values,
with $n_{\text M}:=n-n_{\text N}-n_{\text D},$ such that those
$\theta_j$-values lie in
the interval $(0,\pi/2)$ or $(\pi/2,\pi).$
Our analysis takes into consideration the special cases
where any of $n_{\text N},$ $n_{\text D},$ and $n_{\text M}$ may be zero or $n.$
In our notation the subscripts M, D, and N refer
to ``mixed," ``Dirichlet," and ``Neumann," respectively.
We assume that
the $\theta_j$-values in (4.2) are ordered in such a way that
the first $n_{\text M}$ values of $\theta_j$ correspond to the mixed conditions, the next
$n_{\text D}$ values correspond to the Dirichlet conditions, and the remaining
$n_{\text N}$ values correspond to the Neumann conditions.

We will provide the
explicit steps to go from any pair of matrices $A$ and $B$ satisfying (1.6)-(1.8)
to the pair $\tilde A$ and $\tilde B$ given in (4.2) and yet still satisfying
(1.6)-(1.8) with $\tilde A$ and $\tilde B$ replacing $A$ and $B,$ respectively, there.
For this, we need some auxiliary results.

Starting with $A$ and $B$ satisfying (1.6)-(1.8), let us define
\begin{equation}E:=(A^\dagger A+B^\dagger B)^{1/2},\label{4.3}\end{equation}
so that $E$ is positive, and hence $E$ is uniquely defined.

{\bf Proposition 4.2:} {\it Let $A$ and $B$
be a pair of matrices satisfying (1.6) and (1.7), and let $E$
be the matrix defined in (4.3). Then:}

\begin{itemize}
\item[(a)] {\it The matrix $E$ is invertible and satisfies}
\begin{equation}E=E^\dagger, \quad E^{-1}(A^\dagger A+B^\dagger B)E^{-1}=I_n.
\label{4.4}\end{equation}

\item[(b)] {\it We have}
\begin{equation}(B\pm iA)E^{-2}(B^\dagger\mp iA^\dagger)=I_n,\label{4.5}\end{equation}
{\it and hence the matrices $(B\pm iA)$ and $(B^\dagger\pm iA^\dagger)$
are all invertible and in fact}
\begin{equation}(B\pm i A)^{-1}=E^{-2}(B^\dagger\mp iA^\dagger).\label{4.6}\end{equation}

\item[(c)] {\it The matrix $U$ defined as}
\begin{equation}U:=(B-iA)E^{-2}(B^\dagger-iA^\dagger),\label{4.7}\end{equation}
{\it is unitary, and hence it satisfies $UU^\dagger=U^\dagger U=I_n.$}

\item[(d)] {\it The matrix $U$ defined in (4.7) can also be written as
\begin{equation}U=(B-iA)(B+iA)^{-1},\label{4.8}\end{equation}
and hence from $U^\dagger=U^{-1}$ it follows that}
$$U^\dagger=(B+iA)(B-iA)^{-1}.$$
\end{itemize}

{\it{Proof:}} The proof of (a) readily follows from (4.3). To prove (b)
we let
$$C:=\left[ \begin{array} {cc} BE^{-1}&AE^{-1}\\
AE^{-1}&-BE^{-1}\end{array}\right],$$
and, by using (1.7) and (4.4), we directly verify that
$C^\dagger C=I_{2n},$ proving
the unitarity of $C.$ We must then
also have $CC^\dagger=I_{2n},$ implying
\begin{equation}AE^{-2}A^\dagger+BE^{-2}B^\dagger=I_n,\quad
BE^{-2}A^\dagger-AE^{-2}B^\dagger=0.\label{4.9}\end{equation}
With the help of (4.9), from the identity
$$(B\pm iA)E^{-2}(B^\dagger\mp iA^\dagger)=
(AE^{-2}A^\dagger+BE^{-2}B^\dagger)\mp i
(BE^{-2}A^\dagger-AE^{-2}B^\dagger),$$
we obtain (4.5) and hence (b) is proved. Let us now turn to
the proof of (c).
With the help of (4.3) and (4.5) we directly verify that
the matrix $U$ defined in (4.7) satisfies $UU^\dagger=I_n,$
and hence (c) is proved. From (4.6) and (4.7) we get
(4.8) and hence (d) is also proved. \qed

{\bf Proposition 4.3:} {\it Let $A$ and $B$ be a pair of matrices satisfying (1.7) and (1.8),
and let $\tilde A$ and $\tilde B$
be the matrix pair appearing
in (4.2). We then have}
\begin{equation}\tilde A=M^\dagger AT_1 MT_2,\quad \tilde B=M^\dagger BT_1 MT_2,\label{4.10}\end{equation}
{\it for some unitary matrix $M$ and for some invertible matrices $T_1$ and $T_2.$}

{\it{Proof:}}
We can diagonalize the unitary matrix $U$ appearing in (4.7) and
(4.8) by using a unitary matrix $M$
so that
\begin{equation}M^\dagger UM=\text{diag}\{e^{2i\zeta_1},\dots,e^{2i\zeta_n}\},\label{4.11}\end{equation}
where the constant parameters satisfy $\zeta_j\in(0,\pi].$
Let us define
\begin{equation}Y:=\text{diag}\{e^{i\zeta_1},\dots,e^{i\zeta_n}\}.\label{4.12}\end{equation}
With the help of a permutation matrix $P,$
we can reorder $\zeta_j$ as $\theta_j$ in the manner
described below (4.2), namely, the first
$n_{\text M}$ values of $\theta_j$ lie in
$(0,\pi/2)\cup(\pi/2,\pi),$ the next $n_{\text D}$ values
of $\theta_j$ are all equal to $\pi,$ and the remaining
$n_{\text N}$ values of $\theta_j$ are all equal to $\pi/2.$
Thus, from (4.11) and (4.12) we obtain
\begin{equation}\begin{cases} M^\dagger UMP=Y^2P=\text{diag}\{e^{2i\theta_1},\dots,e^{2i\theta_n}\},\\
YP=\text{diag}\{e^{i\theta_1},\dots,e^{i\theta_n}\},\\
Y^{-1}P=\text{diag}\{e^{-i\theta_1},\dots,e^{-i\theta_n}\}
.\end{cases}\label{4.13}\end{equation}
On the other hand, from (4.2) we see that
\begin{equation}\tilde B-i\tilde A=\text{diag}\{e^{i\theta_1},\dots,e^{i\theta_n}\}
,\label{4.14}\end{equation}
and hence
\begin{equation}(\tilde B-i\tilde A)^{-1}=\tilde B+i\tilde A=
\text{diag}\{e^{-i\theta_1},\dots,e^{-i\theta_n}\}.\label{4.15}\end{equation}
We are now ready to prove (4.10). The invertibility of
$(B+iA)$ is assured by (4.6)
and hence we have
\begin{equation}(B+iA)(B+iA)^{-1}=I_n.\label{4.16}\end{equation}
 From (4.8) we have
\begin{equation}(B-iA)(B+iA)^{-1}=U.\label{4.17}\end{equation}
Let us premultiply (4.16) and (4.17) by $M^\dagger$ and postmultiply them by
$MY^{-1}P$ in order to obtain
\begin{equation}M^\dagger(B+iA)(B+iA)^{-1}MY^{-1}P=Y^{-1}P,\label{4.18}\end{equation}
\begin{equation}M^\dagger(B-iA)(B+iA)^{-1}MY^{-1}P=M^\dagger UMY^{-1}P,\label{4.19}\end{equation}
where we have used the unitarity property $M^\dagger M=I_n.$
 From (4.11) and (4.12) we know that $M^\dagger UM=Y^2$ and hence
$M^\dagger UMY^{-1}P=YP.$ Thus, we can rewrite (4.19) as
\begin{equation}M^\dagger(B-iA)(B+iA)^{-1}MY^{-1}P=YP.\label{4.20}\end{equation}
On the other hand, from (4.13)-(4.15) we see that
\begin{equation}YP=\tilde B-i\tilde A,\quad Y^{-1}P=\tilde B+i\tilde A.\label{4.21}\end{equation}
Using (4.21) we can rewrite (4.18) and (4.20) as
\begin{equation}M^\dagger(B+iA)(B+iA)^{-1}M(\tilde B+i\tilde A)=\tilde B+i\tilde A,\label{4.22}\end{equation}
\begin{equation}M^\dagger(B-iA)(B+iA)^{-1}M(\tilde B+i\tilde A)=\tilde B-i\tilde A.\label{4.23}\end{equation}
Letting
\begin{equation}T_1:=(B+iA)^{-1},\quad T_2:=\tilde B+i\tilde A,\label{4.24}\end{equation}
because of (4.6) and (4.15) we observe that
$T_1$ and $T_2$ are invertible and in fact
\begin{equation}T_1^{-1}=B+iA,\quad T_2^{-1}=\tilde B-i\tilde A.\label{4.25}\end{equation}
Using (4.24) we can rewrite (4.22) and (4.23), respectively,
as
\begin{equation}M^\dagger(B+iA)T_1MT_2=\tilde B+i\tilde A,\label{4.26}\end{equation}
\begin{equation}M^\dagger(B-iA)T_1MT_2=\tilde B-i\tilde A.\label{4.27}\end{equation}
By subtracting and adding, respectively, from (4.26) and (4.27) we obtain
(4.10). \qed

Because of Propositions~4.1 and 4.3, there is not much loss of
generality in using the special boundary parametrization with $\tilde A$ and $\tilde B$
given in (4.2). The relevant results can then be transformed to obtain
the corresponding results in the parametrization with $A$ and $B$ appearing in (1.6)-(1.8).
Let us use a tilde
to denote the quantities obtained under the transformation
given in Proposition~4.1(c). With the help of (4.1) and (4.10),
we can obtain the corresponding quantities in any
boundary parametrization with $A$ and $B$ satisfying (1.7) and (1.8).
In other words, when we have
\begin{equation}V(x)=M\, \tilde V(x)\,M^\dagger,\quad
A=M \tilde AT_2^{-1} M^\dagger T_1^{-1},\quad B=M \tilde BT_2^{-1} M^\dagger T_1^{-1},\label{4.28}\end{equation}
we then get
\begin{equation}f(k,x)=M \tilde f(k,x)\,M^\dagger,\quad
J(k)=M \tilde J(k)\,T_2^{-1} M^\dagger T_1^{-1},\quad
S(k)=M \tilde S(k)\,M^\dagger,\label{4.29}\end{equation}
where $M$ is the unitary matrix appearing in (4.11)
and $T_1^{-1}$ and $T_2^{-1}$ are the matrices specified in (4.25).

\section{The Jost and scattering matrices with zero potential}

In order to understand the large-$k$ behavior of the Jost matrix $J(k),$
its inverse $J(k)^{-1},$ and the scattering matrix
$S(k),$ we need to understand those behaviors
when the
potential $V$ appearing in (1.1) is identically zero.
In that case, let us use the subscript $0$ to denote the corresponding
quantities and write $J_0(k)$ and $S_0(k)$ for
the Jost and scattering matrices, respectively, corresponding to $V\equiv 0.$

When $V\equiv 0,$ from (2.2) we get $f(k,x)=e^{ikx}I_n,$ and hence (3.2)
and (3.8) yield
\begin{equation}J_0(k)=B-ikA, \quad J_0(k)^{-1}=(B-ikA)^{-1},
\quad S_0(k)=-(B+ikA)(B-ikA)^{-1}.\label{5.1}\end{equation}
In the boundary parametrization with $\tilde A$ and $\tilde B$
in (4.2), the corresponding quantities are given by diagonal
matrices, where we have
\begin{equation}\tilde J_0(k)=\tilde B-ik\tilde A=
\text{diag}\{\cos \theta_1+ik\sin\theta_1 ,\dots,\cos\theta_{n_{\text M}}+ik\sin\theta_{n_{\text M}},-I_{n_{\text D}},ikI_{n_{\text N}}\},
\label{5.2}\end{equation}
\begin{equation}\tilde J_0(k)^{-1}=
\text{diag}\left\{{\displaystyle}\frac{1}{\cos \theta_1+ik\sin\theta_1},\dots,
{\displaystyle}\frac{1}{\cos\theta_{n_{\text M}}+ik\sin\theta_{n_{\text M}}},-I_{n_{\text D}},{\displaystyle}\frac{1}{ik}\,I_{n_{\text N}}\right\},
\label{5.3}\end{equation}
\begin{equation}\tilde S_0(k)=
\text{diag}\left\{{\displaystyle}\frac{-\cos \theta_1+ik\sin\theta_1}{\cos \theta_1+ik\sin\theta_1},\dots,
{\displaystyle}\frac{-\cos\theta_{n_{\text M}}+ik\sin\theta_{n_{\text M}}}{\cos\theta_{n_{\text M}}+ik\sin\theta_{n_{\text M}}},
-I_{n_{\text D}},I_{n_{\text N}}\right\}.
\label{5.4}\end{equation}
Then, from (5.3) and (5.4) we see that, as $k\to\infty$ in ${\bf C},$
\begin{equation}\tilde J_0(k)^{-1}=\text{diag}\{0_{n_{\text M}},-I_{n_{\text D}},0_{n_{\text N}}\}
+{\displaystyle}\frac{1}{ik}\,
\text{diag}\{\csc\theta_1,\dots,\csc\theta_{n_{\text M}},0_{n_{\text D}},I_{n_{\text N}}\}+
O(1/k^2),\label{5.5}\end{equation}
\begin{equation}\tilde S_0(k)=Z_0+{\displaystyle}\frac{2i}{k}\,
Z_1+O(1/k^2),\label{5.6}\end{equation}
where we have defined
\begin{equation}Z_0:=\text{diag}\{I_{n_{\text M}},-I_{n_{\text D}},I_{n_{\text N}}\},\quad
Z_1:=\text{diag}\{\cot\theta_1,\dots,\cot\theta_{n_{\text M}},0_{n_{\text D}},0_{n_{\text N}}\},\label{5.7}\end{equation}
with $0_j$ denoting the $j\times j$ zero matrix.

The results in the following proposition are needed in Section~7.

{\bf Proposition 5.1:} {\it The Jost matrix $J_0(k)$ appearing
in (5.1) is invertible when $k\to\infty$ in ${\bf C}.$ The matrix
$J_0(k)^{-1}$
and the scattering matrix
$S_0(k)$ given in (5.1) satisfy, as $k\to\infty$ in ${\bf C},$}
\begin{equation}J_0(k)^{-1}=T_1MT_2\,\text{diag}\{0_{n_{\text M}},-I_{n_{\text D}},0_{n_{\text N}}\}M^\dagger+O(1/k),
\label{5.8}\end{equation}
\begin{equation}A\,J_0(k)^{-1}={\displaystyle}\frac{1}{ik}\,M \,\text{diag}\{-I_{n_{\text M}},0_{n_{\text D}},-I_{n_{\text N}}\}M^\dagger+O(1/k^2),
\label{5.9}\end{equation}
\begin{equation}S_0(k)=S_0(\infty)+{\displaystyle}\frac{2i}{k}\, M
Z_1
M^\dagger+O(1/k^2),\quad S_0(\infty):=MZ_0
M^\dagger
,\label{5.10}\end{equation}
{\it where $T_1$ and $T_2$ are the matrices in (4.24),
$M$ is the unitary matrix in (4.11),
$A$ and $B$ are the matrices appearing in (1.6)-(1.8),
$\tilde A$ and $\tilde B$ are the matrices defined in (4.2),
and $Z_0$ and $Z_1$ are the matrices defined in (5.7).
Thus, as $k\to\infty$ in ${\bf C}$ we have}
\begin{equation}J_0(k)^{-1}=O(1),\quad AJ_0(k)^{-1}=O(1/k),\quad S_0(k)=S_0(\infty)+O(1/k).\label{ 5.11}\end{equation}

{\it{Proof:}} Exploiting the unitarity properties $M^{-1}=M^\dagger$ and
$M^\dagger M=I_n,$ from (4.28) and (4.29) when $V\equiv 0$ we
get
\begin{equation}J_0(k)^{-1}=T_1 M T_2\, \tilde J_0(k)^{-1}M^\dagger,
\quad A\,J_0(k)^{-1}=M \tilde A \,\tilde J_0(k)^{-1}M^\dagger,
\quad S_0(k)=M\tilde S_0(k) M^\dagger.\label{5.12}\end{equation}
 From (5.3) we conclude that
$\tilde J_0(k)$ is invertible when $k\to\infty$ in ${\bf C},$
and hence the first equality in (5.12) implies that
$J_0(k)^{-1}$
exists when $k\to\infty$ in ${\bf C}.$
Using (4.2), (5.5), and (5.6) in (5.12), we get the expansions
(5.8)-(5.10) as $k\to\infty$ in ${\bf C}.$
\qed

\section{Small-$k$ behavior}

The analysis of (1.1) near and at $k=0$ deserves a separate attention and
we refer the reader to Ref.~\citenum{2} for such an analysis. Under the assumption that
the potential $V$ satisfies (1.3) and (1.5), we recall
certain useful properties that will be needed in establishing Levinson's theorem.

When $k=0$ from (1.1) we obtain the zero-energy matrix Schr\"odinger equation
\begin{equation}\psi''=V(x)\,\psi,\qquad x\in{\bf R}^+.\label{6.1}\end{equation}
It is known \cite{1,2} that the Jost solution $f(k,x)$ to (1.1) appearing
in (2.1) satisfies (6.1) if we replace $k$ by $0$ in $f(k,x),$ and that
$f(0,x)$
satisfies \cite{1,2}
\begin{equation}f(0,x)=I_n+o(1),\quad f'(0,x)=o(1/x),\qquad x\to+\infty.\label{6.2}\end{equation}
It is also known \cite{1,2} that
$g(0,x),$ obtained by replacing $k$ with $0$ in the matrix-valued function $g(k,x)$
appearing in (2.3), is a solution to (6.1) and it satisfies
\begin{equation}g(0,x)=x[I_n+o(1)],\quad g'(0,x)=I_n+o(1),\qquad x\to+\infty.\label{6.3}\end{equation}
Thus, from (6.2) and (6.3) we see that
the combined $2n$ columns of $f(0,x)$ and $g(0,x)$
form a fundamental set of solutions to (6.1),
and hence (2.4) is valid even when $k=0.$ From (6.2) we see that
the $n$ columns of $f(0,x)$ form
$n$ linearly independent solutions to (6.1) that remain
bounded as $x\to+\infty.$ Similarly,
(6.3) indicates that the $n$ columns of $g(0,x)$ form
$n$ linearly independent solutions to (6.1) that become
unbounded as $x\to+\infty.$

On the left-hand side of (2.4)
with $k=0,$ let us use a linear combination of
$n$ columns of the zero-energy regular solution $\varphi(0,x),$
which itself is an $n\times n$ matrix solution to (6.1),
and let us express such a column-vector solution as linear combinations of
the $2n$ combined columns of $f(0,x)$ and $g(0,x),$ i.e. we let
\begin{equation}\varphi(0,x)\,u=f(0,x)\,\xi+g(0,x)\,\eta,\label{6.4}\end{equation}
for some nonzero constant column vector $u\in{\bf C}^n.$
We know that the left-hand side in (6.4) satisfies (1.6) for any
$u\in{\bf C}^n$ because $\varphi(0,x)$ itself satisfies (1.6).
We are interested in knowing how many linearly independent
bounded column-vector solutions to (6.1) we can form by using linear combinations
of $n$ columns of $\varphi(0,x).$ Equivalently, we are interested knowing
how many of the $n$ linearly independent bounded column-vector solutions to (6.1) also
satisfy (1.6).

There are two possibilities for (6.4)
with a nonzero column vector $u\in{\bf C}^n;$ either
$\varphi(0,x)\,u$ is bounded as $x\to+\infty,$ in which case we
must have $\xi\ne 0$ and $\eta=0,$ or
$\varphi(0,x)\,u$ is unbounded as $x\to+\infty,$ in which case we
must have $\eta\ne 0.$ In fact, the related results
are known and quoted from Ref.~\citenum{2} in the following
proposition.

{\bf Proposition 6.1:} {\it Consider the selfadjoint matrix
Schr\"odinger operator with the selfadjoint boundary condition
(1.6)-(1.8) and with the potential $V$ satisfying (1.3) and (1.5). Then, we have}

\begin{itemize}
\item [(a)] {\it  The nonzero column vector
$u\in {\bf C}^n$ is an eigenvector of the zero-energy Jost matrix $J(0)$
with the zero eigenvalue, i.e.
$u\in \text{Ker}\, [J(0)],$ if and only if
$\varphi(0,x)\,u$ is bounded for $x\in{\bf R}^+.$}

\item [(b)] {\it For any column vector
$u$ in $\text{Ker}\, [J(0)]$ there
exists a unique column vector $\xi$ in $\text{Ker}\, [J(0)^\dagger]$
such that}
$$\varphi(0,x)\,u=f(0,x)\,\xi.$$
{\it The map $u\mapsto\xi$ from
$\text{Ker}\, [J(0)]$ to $\text{Ker}\, [J(0)^\dagger]$ is a bijection.}
\end{itemize}

As in Theorem~3.1 let us use $\mu$ to denote the geometric multiplicity of the zero
eigenvalue of $J(0).$  From Proposition~6.1 it follows that
we can form exactly $\mu$ columns by using linear
combinations of $n$ columns of
the zero-energy regular solution $\varphi(0,x)$ in such a way that those
$\mu$ columns form linearly independent solutions to (6.1) and they
remain bounded as $x\to+\infty.$ Furthermore, each of such $\mu$ column-vector
solutions to (6.1) can also be expressed
as a linear combinations of columns of $f(0,x).$
In that sense, the integer $\mu$ indicates the maximal number of
linearly independent bounded solutions to (6.1)
that also satisfy (1.6), and hence $\mu$
acts as a ``degree" of the exceptional case \cite{2} for the Schr\"odinger
equation (1.1) with the boundary condition (1.6)-(1.8).
In the purely generic case, i.e. when $\mu=0,$ the $2n$ combined columns of $\varphi(0,x)$ and
of $f(0,x)$ are all linearly independent. In that case,
each column of $\varphi(0,x)$ can be expressed as a linear
combination of $n$ linearly independent columns of $g(0,x).$
In the purely exceptional case, i.e. when $\mu=n,$
each column of $\varphi(0,x)$ can be expressed as a linear
combination of $n$ linearly independent columns of $f(0,x).$

 From (3.3), we have the following conclusion.

{\bf Corollary 6.2:} {\it Consider the selfadjoint matrix
Schr\"odinger operator with the selfadjoint boundary condition
(1.6)-(1.8) and with the potential $V$ satisfying (1.3) and (1.5).
Then, the determinant of
the Jost matrix
$J(k)$ defined in (3.1) has the small-$k$ behavior}
\begin{equation}\det J(k)=c_1k^\mu[1+o(1)],\qquad k\to 0 \text{ in }{\overline{{\bf C}^+}},\label{ 6.5}\end{equation}
{\it where $\mu$ is the geometric multiplicity
of the zero eigenvalue of the zero-energy
 Jost matrix $J(0)$ and
$c_1$ is a nonzero constant. In fact, the value of $c_1$ is given by}
$$c_1:=(\det P_1)(\det P_2)(\det {\mathcal{A}}_1)(\det {\mathcal{D}}_0),$$
{\it where $P_1$ and $P_2$ are the $n\times n$ permutation matrices
appearing in (3.3) and hence their
determinants are either $1$ or $-1,$ and
${\mathcal{A}}_1$ and ${\mathcal{D}}_0$ are the invertible matrices
appearing in (3.3) and hence
their determinants are nonzero.}

 From (3.9) we see that the zero-energy scattering matrix
$S(0)$ has only two eigenvalues,
namely $+1$ and $-1.$ In the next proposition we prove that
the eigenvalue $+1$ of $S(0)$ has multiplicity (both geometric and algebraic)
equal to $\mu$ and that the eigenvalue $-1$ has multiplicity (both geometric and algebraic)
equal to $n-\mu.$ Thus, the value of $\mu$ is uniquely determined from $S(0)$
alone.

{\bf Proposition 6.3:} {\it Consider the selfadjoint matrix
Schr\"odinger operator with the selfadjoint boundary condition
(1.6)-(1.8) and with the potential $V$ satisfying (1.3) and (1.5). Let $J(k)$ and
$S(k)$ be the corresponding Jost matrix and the scattering matrix defined in
(3.1) and (3.8), respectively. Then:}

\begin{itemize}
\item [(a)] {\it $S(0)$ has two eigenvalues, which are $+1$ and $-1.$}

\item [(b)] {\it The geometric and algebraic multiplicities
of the eigenvalue $+1$ of $S(0)$ are both equal to $\mu,$ which is the
geometric multiplicity of the zero eigenvalue of $J(0).$}

\item [(c)] {\it The geometric and algebraic multiplicities
of the eigenvalue $-1$ of $S(0)$ are both equal to $n-\mu.$}
\end{itemize}

{\it{Proof:}} By (3.9), we see that $S(0)$ is similar to a lower-triangular matrix
whose diagonal entries coincide with the diagonal entries of the diagonal matrix
$\text{diag}\{I_\mu,-I_{n-\mu}\}.$ Since a similarity transformation does not change the
eigenvalues, we conclude that
$S(0)$ has the eigenvalue $+1$ with the algebraic
multiplicity $\mu$ and the eigenvalue $-1$ with the algebraic multiplicity
$n-\mu.$ We thus only need to determine the geometric multiplicity
for each eigenvalue. By (3.11), we know that $S(0)$ is unitary and hence
it can be diagonalized with a unitary matrix $M_1.$
Thus, $S(0)=M_1 DM_1^\dagger$ for
some diagonal matrix $D,$ with
$\mu$ of the diagonal entries being $+1$ and $n-\mu$ of
them being $-1.$ The algebraic and geometric multiplicities of
eigenvalues remain invariant under a similarity transformation with a unitary matrix, and
furthermore the algebraic and geometric multiplicities of
each eigenvalue of a diagonal matrix are equal to each other.
Thus, the geometric
multiplicities of the eigenvalues $+1$ and $-1$ are given by
$\mu$ and $n-\mu,$ respectively.
\qed

\section{Large-$k$ behavior}

In establishing the large-$k$ behavior of various quantities related to (1.1),
it is sufficient for the potential to satisfy (1.2) rather than the stronger
condition (1.5).
The following matrices will be useful in our large-$k$ analysis:
\begin{equation}Q_1:={\displaystyle}\frac{1}{2}\int_0^\infty dy\,V(y),\quad Q_2(k):={\displaystyle}\frac{1}{2}\int_0^\infty dy\,
e^{2iky}V(y),\label{7.1}\end{equation}
\begin{equation}Q_3:={\displaystyle}\frac{1}{4}\int_0^\infty dz\int_0^z dy\,V(z)\,V(y),\quad
Q_4(k):={\displaystyle}\frac{1}{4}\int_0^\infty dz\int_0^z dy\,e^{2ikz}V(z)\,V(y),\label{7.2}\end{equation}
\begin{equation}Q_5(k):={\displaystyle}\frac{1}{4}\int_0^\infty dz\int_0^z dy\,e^{2iky}V(z)\,V(y),\quad
Q_6(k):={\displaystyle}\frac{1}{4}\int_0^\infty dz\int_0^z dy\,e^{2ik(z-y)}V(z)\,V(y).\label{7.3}\end{equation}
We emphasize that $Q_1$ and $Q_3$ are independent of $k$ while the remaining four matrices
are functions of $k.$

{\bf Proposition 7.1:} {\it Assume that $V$ in (1.1)
belongs to $L^1({\bf R}^+).$ Then, $Q_1$ and $Q_3$ are well defined constant
matrices. Furthermore, each of
the four matrix quantities $Q_2(k),$ $Q_4(k),$ $Q_5(k),$ and $Q_6(k)$
is well defined for $k\in{\overline{{\bf C}^+}},$ and each one of them
has the behavior of $o(1)$ as $k\to\infty$
in ${\overline{{\bf C}^+}}.$}

{\it{Proof:}} The integrals defining $Q_1$ and $Q_3$ exist because
$V\in L^1({\bf R}^+).$ For $k\in{\overline{{\bf C}^+}},$ the coefficients of $2ik$ in the exponents
appearing in $Q_2(k),$ $Q_4(k),$ $Q_5(k),$ and $Q_6(k)$ are all nonnegative, and hence
each of those exponential terms are bounded by one in absolute value.
Furthermore, $V\in L^1({\bf R}^+),$ and hence we have the estimate
$$\left\|\int_0^x dz\,V(z)\right\|\le \int_0^x dz\,||V(z)||\le \int_0^\infty dz\,||V(z)||,
$$
where the integral on the right-hand side converges. Thus, we can establish that the
integrals used in defining $Q_2(k),$ $Q_4(k),$ $Q_5(k),$ $Q_6(k)$ all
exist for $k\in{\overline{{\bf C}^+}}.$ Finally, the behavior of $o(1)$ as $k\to\infty$
in ${\overline{{\bf C}^+}}$ for $Q_2(k),$ $Q_4(k),$ $Q_5(k),$ $Q_6(k)$ is established with
the help of the Riemann-Lebesgue lemma on the appropriate integrals. \qed

{\bf Proposition 7.2:} {\it
Consider the matrix
Schr\"odinger equation (1.1) with the selfadjoint
potential $V$ satisfying (1.2) and (1.3).
Let $f(k,x)$ be the corresponding Jost solution satisfying (2.1).
Then, we have as $k\to\infty$ in ${\overline{{\bf C}^+}}$}
\begin{equation}f(-k^*,0)^\dagger=I_n+{\displaystyle}\frac{1}{ik}\left[-Q_1+Q_2(k)\right]
+{\displaystyle}\frac{1}{k^2}\left[-Q_3-Q_4(k)+Q_5(k)+Q_6(k)\right]+O(1/k^3),\label{7.4}\end{equation}
\begin{equation}f'(-k^*,0)^\dagger=ik I_n-Q_1-Q_2(k)+{\displaystyle}\frac{1}{ik}\left[Q_3-Q_4(k)+Q_5(k)-Q_6(k)\right]+O(1/k^2),
\label{7.5}\end{equation}
{\it where $Q_1,$ $Q_2(k),$ $Q_3,$ $Q_4(k),$ $Q_5(k),$ $Q_6(k)$ are the matrices defined
in (7.1)-(7.3) with the properties outlined in Proposition~7.1.}

{\it{Proof:}} Writing the Jost solution $f(k,x)$ in terms of
$m(k,x):=e^{-ikx}f(k,x),$ from (2.2) we obtain the integral relations
\begin{equation}m(k,x)=I_n+{\displaystyle}\frac{1}{2ik}\int_x^\infty
dy\,\left[e^{2ik(y-x)}-1\right]V(y)\,m(k,y),\label{7.6}\end{equation}
\begin{equation}m'(k,x)=-\int_x^\infty dy\,e^{2ik(y-x)}\,V(y)\,m(k,y).\label{7.7}\end{equation}
Iterating (7.6) and (7.7), for $k\to\infty$ in ${\overline{{\bf C}^+}}$ we obtain
\begin{equation}\begin{aligned}
m(k,x)=&I_n+{\displaystyle}\frac{1}{2ik}\int_x^\infty
dy\,\left[e^{2ik(y-x)}-1\right]V(y)\\
&+{\displaystyle}\frac{1}{(2ik)^2}\int_x^\infty
dy\,\left[e^{2ik(y-x)}-1\right]V(y)\int_y^\infty dz\,
\left[e^{2ik(y-z)}-1\right]V(z)+O(1/k^3),\end{aligned}\label{7.8}\end{equation}
\begin{equation}\begin{aligned}
m'(k,x)=&-\int_x^\infty dy\,e^{2ik(y-x)}\,V(y)
\\
&-
{\displaystyle}\frac{1}{2ik}\int_x^\infty dy\,e^{2ik(y-x)}\,V(y)
\int_y^\infty dz\,
\left[e^{2ik(y-z)}-1\right]V(z)+O(1/k^2).\end{aligned}\label{7.9}\end{equation}
We evaluate (7.8) and (7.9) at $x=0,$ and we
rewrite the double integrals in them by changing the order of
integration.
Then, with the help of
\begin{equation}f(k,0)=m(k,0),\quad f'(k,0)=ik\,m(k,0)+m'(k,0),\label{7.10}\end{equation}
we obtain the expansions for
$f(k,0)$ and $f'(k,0)$ as $k\to\infty$ in ${\overline{{\bf C}^+}}.$
Finally, by replacing $k$ by $-k^*,$ taking the adjoints, and
using $\left[V(y)\,V(z)\right]^\dagger=V(z)\,V(y),$
with the help of (7.8)-(7.10) we obtain (7.4) and (7.5). \qed

{\bf Proposition 7.3:} {\it
Consider
the Schr\"odinger operator corresponding to
(1.1) with the selfadjoint boundary condition
(1.6)-(1.8) and with the potential
$V$ satisfying (1.2) and (1.3). Let $J(k)$ be the corresponding Jost matrix defined
in (3.1). Then, we have as $k\to\infty$ in ${\overline{{\bf C}^+}}$}
\begin{equation}J(k)=-ikA+B+[Q_1+Q_2(k)]A+{\displaystyle}\frac{1}{ik}P(k)+O(1/k^2),\label{7.11}\end{equation}
{\it where we have defined}
$$P(k):=[-Q_1+Q_2(k)]B+[-Q_3+Q_4(k)-Q_5(k)+Q_6(k)]A,$$
{\it with $A$ and $B$ being the matrices appearing in (1.6)-(1.8),
and $Q_1,$ $Q_2(k),$ $Q_3,$ $Q_4(k),$ $Q_5(k),$
$Q_6(k)$ being the matrices
in (7.1)-(7.3).}

{\it{Proof:}} Using (7.4) and (7.5) in (3.2) we obtain (7.11). \qed

{\bf Proposition 7.4:} {\it Consider
the Schr\"odinger operator corresponding to
(1.1) with the selfadjoint boundary condition
(1.6)-(1.8) and with the potential
$V$ satisfying (1.2) and (1.3). Let $J(k)$ and $J_0(k)$
be the corresponding Jost matrices appearing in (3.1) and (5.1), respectively,
$Q_1$ and $Q_2(k)$ be the quantities in (7.1), and
$S_0(\infty)$ be the constant matrix defined in (5.10).
Then, as $k\to\infty$ in ${\overline{{\bf C}^+}}$ we have}
\begin{equation}J(k)\,J_0(k)^{-1}=I_n-{\displaystyle}\frac{1}{ik}\left[
Q_1+Q_2(k)\,S_0(\infty)\right]+O(1/k^2),
\label{7.12}\end{equation}
\begin{equation}J_0(k)\,J(k)^{-1}=I_n+{\displaystyle}\frac{1}{ik}\left[
Q_1+Q_2(k)\,S_0(\infty)\right]+O(1/k^2).
\label{7.13}\end{equation}

{\it{Proof:}} By replacing $B-ikA$ by $J_0(k),$ as given in (5.1), and
replacing $B+ikA$ by $J_0(-k),$ from (7.11) we get,
as $k\to\infty$ in ${\overline{{\bf C}^+}},$
\begin{equation}J(k)=J_0(k)-{\displaystyle}\frac{1}{ik}\,Q_1J_0(k)+{\displaystyle}\frac{1}{ik}\,Q_2(k)\,J_0(-k)
+{\displaystyle}\frac{1}{ik}[-Q_3+Q_4(k)-Q_5(k)+Q_6(k)]A+O(1/k^2)
.\label{7.14}\end{equation}
The invertibility of $J_0(k)$ as $k\to\infty$ in ${\overline{{\bf C}^+}}$ is assured
by Proposition~5.1.
Let us multiply (7.14) on the right by $J_0(k)^{-1}$ and use
$J_0(-k)\,J_0(k)^{-1}=-S_0(k),$ which follows from (3.8). We then obtain,
as $k\to\infty$ in ${\overline{{\bf C}^+}},$
\begin{equation}\begin{aligned} J(k)\,J_0(k)^{-1}=&I_n-{\displaystyle}\frac{1}{ik}\,Q_1-{\displaystyle}\frac{1}{ik}\,Q_2(k)\,S_0(k)
\\
&+{\displaystyle}\frac{1}{ik}[-Q_3+Q_4(k)-Q_5(k)+Q_6(k)]A\,J_0(k)^{-1}
+O(1/k^2)\,J_0(k)^{-1}
.\end{aligned}\label{7.15}\end{equation}
Using (5.11) in (7.15), we get
$$J(k)\,J_0(k)^{-1}=I_n-{\displaystyle}\frac{1}{ik}\,Q_1-{\displaystyle}\frac{1}{ik}\,Q_2(k)\,S_0(k)
+O(1/k^2),\qquad k\to\infty\text { in }{\overline{{\bf C}^+}},$$
which is also equivalent to (7.12) because of the third
estimate in (5.11).
The expansion in (7.13) is obtained by changing the signs
of the $O(1/k)$-terms in (7.12). \qed

{\bf Proposition 7.5:} {\it Consider
the Schr\"odinger operator corresponding to
(1.1) with the selfadjoint boundary condition
(1.6)-(1.8) and with the potential
$V$ satisfying (1.2) and (1.3). Then, the corresponding Jost matrix $J(k)$
defined in (3.1)
satisfies}
\begin{equation}J(k)=J_0(k)[I_n+O(1/k)],\qquad k\to \infty\text{ in }{\overline{{\bf C}^+}},
\label{7.16}\end{equation}
\begin{equation}\det J(k)=c_2 k^{n_{\text M}+n_{\text N}}[1+O(1/k)],\qquad k\to \infty\text{ in }{\overline{{\bf C}^+}},
\label{7.17}\end{equation}
{\it where $J_0(k)$ is the matrix in (5.1), $c_2$ is a nonzero constant,
and $n_{\text M}$ and $n_{\text N}$ are the nonnegative integers defined
after (4.2) and appearing in (5.2)-(5.4).}

{\it{Proof:}} Note that (7.16) is apparent from (7.15). With the help of
the first equality in
(5.12) we get
\begin{equation}J_0(k)=M\tilde J_0(k)\,T_2^{-1}M^\dagger T_1^{-1},\label{7.18}\end{equation}
where $T_1$ and $T_2$ are the invertible matrices appearing
in (4.24) and (4.25),
$M$ is the unitary matrix appearing in (4.11), and
$\tilde J_0(k)$ is the matrix in (5.2). Using (5.2) and
(7.18) we obtain
\begin{equation}\det J_0(k)=c_2 k^{n_{\text M}+n_{\text N}}[1+O(1/k)],\qquad
k\to \infty\text{ in }{\overline{{\bf C}^+}},\label{7.19}\end{equation} where we have defined
$$c_2:={\displaystyle}\frac{(-1)^{n_{\text D}}(i)^{n_{\text M}+n_{\text N}}}{\det[T_1T_2]}\prod_{j=1}^{n_{\text M}}\sin \theta_j.
$$
Note that $c_2$ is well defined and nonzero because
$T_1$ and $T_2$ are invertible matrices and $\sin\theta_j\ne 0$
for $j=1,\dots,n_{\text M}.$ The latter follows from the fact that
those $\theta_j$ all lie in $(0,\pi/2)\cup(\pi/2,\pi),$
as stated above (4.13). From (7.16) and (7.19) we then get (7.17).
\qed

Next we present the large-$k$ asymptotics of the scattering matrix.

{\bf Theorem 7.6:} {\it Consider
the Schr\"odinger operator corresponding to
(1.1) with the selfadjoint boundary condition
(1.6)-(1.8) and with the potential
$V$ satisfying (1.2) and (1.3). Then the corresponding scattering matrix $S(k)$
defined in (3.8)
satisfies:}
\begin{equation}S(k)=S_0(\infty)+{\displaystyle}\frac{G(k)}{ik}+O(1/k^2),\qquad k\to\pm\infty,
\label{7.20}\end{equation}
{\it where $G(k)$ is the matrix defined as}
$$G(k):=-2MZ_1 M^\dagger+Q_1 S_0(\infty)+S_0(\infty)\,Q_1+S_0(\infty)Q_2(k)S_0(\infty)+Q_2(-k),
$$
{\it with $M$ being the unitary matrix
in (4.11), $Z_1$ the matrix in (5.7), $S_0(\infty)$ the matrix in (5.10),
and $Q_1$ and $Q_2(k)$ the matrices in (7.1).}

{\it{Proof:}} With the help of (3.8) we see that
\begin{equation}S(k)=J(-k)\,J_0(-k)^{-1}S_0(k)J_0(k)\,J(k)^{-1},\label{7.21}\end{equation}
where $J_0(k)$ and $S_0(k)$ are the matrices defined in (5.1)
and $J(k)$ is the Jost matrix appearing in (3.1).
Using (7.12) and (7.13) in (7.21), with the help of
the identity $S_0(-k)S_0(k)=I_n,$ which follows from (3.11),
we obtain
\begin{equation}S(k)=S_0(k)+{\displaystyle}\frac{H(k)}{ik}+O(1/k^2),\qquad k\to\pm\infty,\label{7.22}\end{equation}
where we have defined
\begin{equation}H(k):=Q_1S_0(k)+S_0(k)Q_1+S_0(k)Q_2(k)S_0(k)+Q_2(-k).\label{7.23}\end{equation}
Finally, using (5.10) in (7.22) and (7.23), we obtain (7.20). \qed

\section{Bound states}

A bound state for the Schr\"odinger equation (1.1) with the boundary condition
(1.6)-(1.8) at $x=0$ corresponds to a square-integrable column-vector solution
satisfying (1.6). Because of (1.3) and (1.6)-(1.8), the corresponding
Schr\"odinger operator is selfadjoint, and hence a bound state must occur
at a real value of $k^2.$ From (2.1), (2.3), and (2.4), we see that we cannot
have any bound states at any positive values of $k^2$
because none of the combined $2n$ linearly independent
columns of $f(k,x)$ and $g(k,x)$ can be square integrable on $x\in{\bf R}^+$
when $k$ is real.
Similarly, from (6.2) and (6.3) we
see that none of the combined $2n$ linearly independent
columns of $f(0,x)$ and $g(0,x)$ can be square integrable on $x\in{\bf R}^+,$
and since (2.4) also holds at $k=0,$ we can conclude that there cannot be a bound state
when $k=0.$ Thus, a bound state, if it exists, can only occur when $k^2<0,$ which
corresponds to a value of $k$ on the positive imaginary axis in ${\bf C}.$

Let us assume that $k=i\kappa$ for some positive $\kappa$ corresponds
to a bound state, and let $\omega(i\kappa,x)$
be a square-integrable column-vector solution satisfying (1.6). Then, (2.4) must hold
at $k=i\kappa$ with $\eta=0$ and for some nonzero column vector $\xi\in{\bf C}^n,$ yielding
\begin{equation}\omega(i\kappa,x)=f(i\kappa,x)\xi.\label{8.1}\end{equation}
Let us show that $\xi$ must belong to the kernel of
$J(i\kappa)^\dagger$ because $\omega(i\kappa,x)$ must satisfy (1.6).
Note that, from (1.6) and (8.1) we get
\begin{equation}-B^\dagger f(i\kappa,0)\xi+A^\dagger f'(i\kappa,0)\xi=0,\label{8.2}\end{equation}
which is equivalent to
\begin{equation}\left[-\varphi'(i\kappa,0)^\dagger f(i\kappa,0)+\varphi(i\kappa,0)^\dagger
f'(i\kappa,0)\right]\xi=0,\label{8.3}\end{equation}
or equivalently, in terms of the Wronskian, (8.3) can be written as
\begin{equation}[f(i\kappa,x)^\dagger,\varphi(i\kappa,x)]^\dagger \xi=0.\label{8.4}\end{equation}
Comparing (3.1) and (8.4), we see that (8.4) is equivalent to
\begin{equation}J(i\kappa)^\dagger \xi=0,\label{8.5}\end{equation}
and hence $\xi$ belongs to $\text{Ker}[J(i\kappa)^\dagger].$ Thus,
the determinant of $J(i\kappa)$ must be zero.

Conversely, consider any column vector of the form
$f(i\kappa,x)\,\xi,$ where $k=i\kappa$ corresponds to a zero
of $\det J(k) $ on the positive imaginary axis and
$\xi\in{\bf C}^n$ is a nonzero column vector belonging to
$\text{Ker}[J(i\kappa)^\dagger].$ Then, $f(i\kappa,x)\,\xi$
must be a bound-state column-vector solution to the corresponding Schr\"odinger operator.
To verify this, we must prove that
$f(i\kappa,x)\,\xi$ is a solution to (1.1), it is
square integrable on $x\in{\bf R}^+,$ and it satisfies the boundary condition (1.6)-(1.8).
It is clearly a solution to (1.1) because $f(k,x)$ is a $n\times n$ matrix solution
to (1.1). It is square integrable because $f(k,x)$ exponentially decays to zero
for each $k\in{{\bf C}^+}$
as $x\to+\infty,$ as apparent from (2.1). Finally, it satisfies the boundary
condition (1.6) because
$$-B^\dagger f(i\kappa,0)\xi+A^\dagger f'(i\kappa,0)\xi=J(i\kappa)^\dagger \xi=0,$$
as seen from (8.2)-(8.5). The multiplicity of the bound state at $k=i\kappa$
is equal to the dimension of the kernel of $J(i\kappa)^\dagger,$ which is also
equal to the dimension of the kernel of $J(i\kappa).$

Let us now show that a bound-state column-vector solution at $k=i\kappa$
must have the form $\varphi(i\kappa,x)\alpha$ for some constant nonzero column vector
$\alpha\in{\bf C}^n$ that belongs to the kernel of $J(i\kappa)$ in such a way that
\begin{equation}\varphi(i\kappa,x)\alpha=f(i\kappa,x)\beta,\label{8.6}\end{equation}
where $\beta\in \text{Ker}[J(i\kappa)^\dagger].$
In other words, we must show that $\varphi(i\kappa,x)\alpha$ with
$\alpha\in\text{Ker}[J(i\kappa)]$ satisfies
(1.6), is square integrable, and satisfies (8.6) for some column vector
$\beta\in\text{Ker}[J(i\kappa)^\dagger].$
Note that (1.6) is satisfied because with the help of (1.7) and (2.5) we get
$$-B^\dagger \varphi(i\kappa,0)\alpha+A^\dagger \varphi'(i\kappa,0)\alpha
=(-B^\dagger A+A^\dagger B)\alpha=0.$$
Next, let us show that $\varphi(i\kappa,x)\alpha$ exponentially decays to zero
as $x\to+\infty,$ and hence it is square integrable on $x\in{\bf R}^+.$ From (2.4)
we see that $\varphi(i\kappa,x)\alpha$ can be written as a linear combination of
the $2n$ linearly independent columns of $f(i\kappa,x)$ and $g(i\kappa,x),$ i.e.
there exist some constant column vectors $\beta$ and $\gamma$ in
${\bf C}^n$ such that
\begin{equation}\varphi(i\kappa,x)\,\alpha=f(i\kappa,x)\,\beta+g(i\kappa,x)\,\gamma,\qquad
x\in{\bf R}^+.\label{8.7}\end{equation}
Let us now evaluate the Wronskian-related quantity
$[f(i\kappa,x)^\dagger;\varphi(i\kappa,x)]\alpha$ using (8.7). With the help of
(3.1) we get
\begin{equation}[f(i\kappa,x)^\dagger;\varphi(i\kappa,x)]\alpha=J(i\kappa)\alpha=0,\label{8.8}\end{equation}
because $\alpha\in\text{Ker}[J(i\kappa)].$
On the other hand, using (8.7) we get
\begin{equation}\begin{aligned}
[f(i\kappa,x)^\dagger;\varphi(i\kappa,x)]\alpha&=
[f(i\kappa,x)^\dagger;f(i\kappa,x)\,\beta+g(i\kappa,x)\,\gamma]\\
&=[f(i\kappa,x)^\dagger;f(i\kappa,x)]\,\beta+[f(i\kappa,x)^\dagger;g(i\kappa,x)]\,\gamma\\
&=[f(i\kappa,x)^\dagger;g(i\kappa,x)]\,\gamma\\
&=2\kappa \gamma,
\end{aligned}\label{8.9}\end{equation}
where we have used
(2.1) and (2.3) to evaluate
the relevant Wronskians.
Comparing (8.8) and (8.9) we see that $\gamma=0$ and hence
(8.6) is satisfied for some nonzero column vector $\beta$ in ${\bf C}^n.$
Because of (2.1), from (8.6) we conclude that $\varphi(i\kappa,x)\alpha$
decays exponentially to zero as $x\to+\infty$ and hence
it is square integrable. Note that, $\beta$ must belong
to $\text{Ker}[J(i\kappa)^\dagger]$ as a result of our earlier argument
that if $f(i\kappa,x)\beta$ is a bound state then
$\beta$ must belong to the kernel of $J(i\kappa)^\dagger.$

Let us emphasize that (8.6) establishes a bijection $\alpha\mapsto \beta$
between $\text{Ker}[J(i\kappa)]$ and $\text{Ker}[J(i\kappa)^\dagger]$
for any $k=i\kappa$ that is a zero of $\det J(k) $ on the positive imaginary axis.
Since $f(i\kappa,x)\xi$ corresponds to a bound state with
$\xi\in\text{Ker}[J(i\kappa)^\dagger],$ from (2.1) we conclude that
there are as many linearly independent bound states at $k=i\kappa$ as the
dimension of $\text{Ker}[J(i\kappa)^\dagger].$ Since that
is also equal to the dimension of $\text{Ker}[J(i\kappa)],$
we can say that the multiplicity of the bound state at $k=i\kappa$ is given by
the dimension of $\text{Ker}[J(i\kappa)].$ Let us use
$m_{\kappa}$ to denote the multiplicity of the bound state at $k=i\kappa.$
We thus have
\begin{equation}m_{\kappa}=\text{dim}\,\text{Ker}[J(i\kappa)].\label{8.10}\end{equation}
Note that $1\le m_\kappa\le n$ because the dimension of
$\text{Ker}[J(i\kappa)]$ cannot exceed $n$
for the corresponding $n\times n$ matrix $J(k).$

We summarize the above observations on bound states in the following theorem.

{\bf Theorem 8.1:} {\it Consider the selfadjoint matrix
Schr\"odinger operator with the selfadjoint boundary condition
(1.6)-(1.8) and with the potential $V$ satisfying (1.3) and (1.5).
Let $f(k,x),$ $\varphi(k,x),$ and $J(k)$
be the corresponding Jost solution, the regular solution, and
the Jost matrix, appearing in (2.1), (2.5), and (3.1), respectively.
Then:}

\begin{itemize}
\item[(a)] {\it We have a bound state at $k=i\kappa$ for some positive $\kappa$ if
and only if $\text{Ker}[J(i\kappa)]$ is nontrivial
or equivalently if and only if $\text{det}[J(i\kappa)]=0.$}

\item[(b)] {\it The multiplicity $m_\kappa$ of the bound state at
$k=i\kappa$ is finite, and in fact it is equal to the dimension of $\text{Ker}[J(i\kappa)].$}

\item[(c)] {\it A bound-state column-vector solution to (1.1) at $k=i\kappa$
must be equal
to $f(i\kappa,x)\,\beta$ for some nonzero column vector
$\beta\in\text{Ker}[J(i\kappa)^\dagger].$
Similarly, a bound-state column-vector solution to (1.1) at $k=i\kappa$
must be equal
to $\varphi(i\kappa,x)\,\alpha$ for some nonzero column
vector $\alpha\in\text{Ker}[J(i\kappa)].$}

\item[(d)] {\it If $k=i\kappa$ corresponds to a bound state, then
there is a bijection $\alpha\mapsto \beta$
between
$\text{Ker}[J(i\kappa)]$ and $\text{Ker}[J(i\kappa)^\dagger]$ in such a way
that $\varphi(i\kappa,x)\,\alpha=f(i\kappa,x)\,\beta.$}
\end{itemize}

We will next analyze the behaviors of the Jost matrix $J(k)$ and of its inverse at a bound state
$k=i\kappa.$ One of our goals is to prove that the multiplicity $m_\kappa$ of the
bound state is equal to the multiplicity of the zero of $\det J(k)$ at $k=i\kappa.$
We will use an overdot to indicate the derivative with respect to $k.$

{\bf Proposition 8.2:} {\it Consider the selfadjoint matrix
Schr\"odinger operator with the selfadjoint boundary condition
(1.6)-(1.8) and with the potential $V$ satisfying (1.3) and (1.5).
Let $f(k,x)$ and $J(k)$
be the corresponding Jost solution and the Jost matrix,
appearing in (2.1) and (3.2), respectively.
Assume that there is a bound state at $k=i\kappa$ for some
positive $\kappa.$ Then,
for each fixed $x\in{\bf R}^+$ we have}
\begin{equation}f(-k^*,x)^\dagger\big|_{k=i\kappa}=f(i\kappa,x)^\dagger,
\quad
{\displaystyle}\frac{d f(-k^*,x)^\dagger}{dk}\bigg|_{k=i\kappa}=-\dot f(i\kappa,x)^\dagger,
\label{8.11}\end{equation}
\begin{equation}{\displaystyle}\frac{d f'(-k^*,x)^\dagger}{dk}\bigg|_{k=i\kappa}=-\dot f'(i\kappa,x)^\dagger,
\quad
{\displaystyle}\frac{d f''(-k^*,x)^\dagger}{dk}\bigg|_{k=i\kappa}=-\dot f''(i\kappa,x)^\dagger
,
\label{8.12}\end{equation}
\begin{equation}\dot J(i\kappa)=\dot f'(i\kappa)^\dagger A-\dot f(i\kappa,0)^\dagger B.\label{8.13}\end{equation}

{\it{Proof:}} As stated in Section~2, $f(k,x)$ is analytic in
$k\in{{\bf C}^+}$ for each fixed $x\in{\bf R}^+.$ Thus, we have the Taylor series expansion
\begin{equation}f(k,x)=f(i\kappa,x)+(k-i\kappa)\,\dot f(i\kappa,x)+O((k-i\kappa)^2)),
\qquad k\to i\kappa.
\label{8.14}\end{equation}
Replacing $k$ by $-k^*$ in (8.14) and by taking the adjoint of both sides of the
resulting expansion, we get
\begin{equation}f(-k^*,x)^\dagger=f(i\kappa,x)^\dagger-(k-i\kappa)\,\dot f(i\kappa,x)^\dagger+O((k-i\kappa)^2)),
\qquad k\to i\kappa.\label{8.15}\end{equation}
The first and second terms in the expansion on the right-hand side
in (8.15) yield the equalities in (8.11).
The equalities in (8.12) are established
in a similar manner by exploiting the
analyticity of $f'(k,x)$ and $f''(k,x)$ in $k\in{{\bf C}^+}$ for
each fixed $x\in{\bf R}^+.$ By taking the $k$-derivative
of both sides of (3.2) and using the second equality in (8.11) and the
first equality in (8.12), we obtain (8.13). \qed

{\bf Theorem 8.3:} {\it Consider the selfadjoint matrix
Schr\"odinger operator with the selfadjoint boundary condition
(1.6)-(1.8) and with the potential $V$ satisfying (1.3) and (1.5).
Let $f(k,x),$ $\varphi(k,x),$ and $J(k)$
be the corresponding Jost solution, the regular solution, and
the Jost matrix, appearing in (2.1), (2.5), and (3.1), respectively.
Assume that there is a bound state at $k=i\kappa$
for some
positive $\kappa.$
For any constant column vector
$\alpha\in \text{Ker}[J(i\kappa)],$ let $\beta$ be the corresponding unique
constant column vector indicated in Theorem~8.1(d).
Then:}
\begin{equation}i\beta^\dagger \dot J(i\kappa)\alpha=
2\kappa\int_0^\infty dx\,[\varphi(i\kappa,x)\alpha]^\dagger
[\varphi(i\kappa,x)\alpha],\label{8.16}\end{equation}
{\it and hence $\beta^\dagger \dot J(i\kappa)\alpha\ne 0$
unless $\alpha=0.$}

{\it{Proof:}} The Jost solution $f(k,x)$ satisfies (1.1) and hence
\begin{equation}f''(k,x)+k^2 f(k,x)=V(x)\,f(k,x).\label{8.17}\end{equation}
By taking the $k$-derivative, from (8.17) we get
\begin{equation}\dot f''(k,x)+k^2 \dot
f(k,x)+2k f(k,x)=V(x)\,\dot f(k,x),\label{8.18}\end{equation}
and by replacing $k$ by $-k^*$ in (8.17)
and then taking the adjoint we get
\begin{equation}f''(-k^*,x)^\dagger+k^2 f(-k^*,x)^\dagger=f(-k^*,x)^\dagger
V(x),\label{8.19}\end{equation}
where we have used the selfadjointness of $V$ given in (1.3).
Evaluating (8.18) and (8.19) at $k=i\kappa,$ with the help of (8.11), we get
\begin{equation}\dot f''(i\kappa,x)-\kappa^2 \dot
f(i\kappa,x)+2i\kappa f(i\kappa,x)=V(x)\,\dot f(i\kappa,x),\label{8.20}\end{equation}
\begin{equation}f''(i\kappa,x)^\dagger-\kappa^2 f(i\kappa,x)^\dagger=f(i\kappa,x)^\dagger
V(x).\label{8.21}\end{equation}
Premultiplying (8.20) by $f(i\kappa,x)^\dagger$ and postmultiplying
(8.21) by $\dot f(i\kappa,x)$
and taking the difference of the resulting equations, we obtain
\begin{equation}{\displaystyle}\frac{d}{dx}\left[f(i\kappa,x)^\dagger
\dot f'(i\kappa,x)-
f'(i\kappa,x)^\dagger
\dot f(i\kappa,x)\right]=-2i\kappa f(i\kappa,x)^\dagger f(i\kappa,x)
.\label{8.22}\end{equation}
Premultiplying (8.22) by $\beta^\dagger$ and postmultiplying it
by $\beta,$ we integrate the resulting equation over
$x\in{\bf R}^+.$ We then get
\begin{equation}-
\beta^\dagger f(i\kappa,0)^\dagger
\dot f'(i\kappa,0)\beta
+\beta^\dagger f'(i\kappa,0)^\dagger
\dot f(i\kappa,0)\beta=
-2i\kappa\int_0^\infty dx\,[f(i\kappa,x)\beta]^\dagger
[f(i\kappa,x)\beta],\label{8.23}\end{equation}
where we have used Theorem~8.1(c) with the fact that
$f(i\kappa,x)\beta$ is a bound-state column-vector solution and hence it is
square integrable, and we have also used the fact that
the quantity inside the brackets in (8.22) vanishes as
$x\to+\infty.$ The latter property
is a consequence of the exponential decay
to zero of $f(i\kappa,x)\beta$ and
of $f'(i\kappa,x)\beta$ and can be established
with the help of (2.1) at $k=i\kappa.$ Multiplying (8.23) on both sides
by $i$ and using (8.6) we get
\begin{equation}-i\left[
\alpha^\dagger \varphi(i\kappa,0)^\dagger
\dot f'(i\kappa,0)\beta
-\alpha^\dagger \varphi'(i\kappa,0)^\dagger
\dot f(i\kappa,0)\beta\right]=
2\kappa\int_0^\infty dx\,[\varphi(i\kappa,x)\alpha]^\dagger
[\varphi(i\kappa,x)\alpha].\label{8.24}\end{equation}
Using (2.5) and by taking the adjoint of both sides in (8.24) we get
\begin{equation}i\beta^\dagger \left[
\dot f'(i\kappa,0)^\dagger A-
\dot f(i\kappa,0)^\dagger B
\right]\alpha=
2\kappa\int_0^\infty dx\,[\varphi(i\kappa,x)\alpha]^\dagger
[\varphi(i\kappa,x)\alpha].\label{8.25}\end{equation}
Comparing the left-hand side of (8.25) with (8.13), we obtain (8.16).
Finally, since $\kappa>0$ we see that the right-hand side
in (8.16) is positive if $\alpha\ne 0$ and is equal
to zero if $\alpha=0.$ Thus, from the left-hand side of (8.16) we conclude that
$\beta^\dagger\dot J(i\kappa)\alpha=0$ only when $\alpha=0.$
\qed

{\bf Theorem 8.4:} {\it Consider the selfadjoint matrix
Schr\"odinger operator with the selfadjoint boundary condition
(1.6)-(1.8) and with the potential $V$ satisfying (1.3) and (1.5).
Let $f(k,x),$ $\varphi(k,x),$ and $J(k)$
be the corresponding Jost solution, the regular solution, and
the Jost matrix, appearing in (2.1), (2.5), and (3.1), respectively.
Assume that there is a bound state at $k=i\kappa$ for some
positive $\kappa.$ Then,
$J(k)^{-1}$ has a simple pole at $k=i\kappa.$}

{\it{Proof:}} By Theorem~8.1(a), the determinant of $J(k)$
vanishes at $k=i\kappa,$ and hence Theorem~3.1(a) implies that
$J(k)^{-1}$ is analytic in a deleted neighborhood of
$k=i\kappa$ with a pole of some finite order $p$ at $k=i\kappa.$
If the pole
at $k=i\kappa$ were not
simple, then for
$p\ge 2,$ in some neighborhood of $k=i\kappa$ we would have the
expansions
\begin{equation}J(k)=J(i\kappa)+(k-i\kappa)\,\dot J(i\kappa)+O((k-i\kappa)^2),\label{8.26}\end{equation}
\begin{equation}J(k)^{-1}={\displaystyle}\frac{N_{-p}}{(k-i\kappa)^p}+
{\displaystyle}\frac{N_{-p+1}}{(k-i\kappa)^{p-1}}+\dots+
{\displaystyle}\frac{N_{-1}}{k-i\kappa}+N_0
+(k-i\kappa)\,N_1+O((k-i\kappa)^2).\label{8.27}\end{equation}
Using (8.26) and (8.27) in
$J(k)J(k)^{-1}=I_n,$ we would obtain
\begin{equation}J(i\kappa)\,N_{-p}=0,\quad J(i\kappa)\,N_{-p+1}+\dot J(i\kappa)\,N_{-p}=0.\label{ 8.28}\end{equation}
 From the first equation in (8.28) we see that each column of
$N_{-p}$ would have to belong to $\text{Ker}[J(i\kappa)].$ For each
nonzero column of $N_{-p},$ by denoting that nonzero column
with $\alpha$ as in Theorem~8.3, from (8.28) we would get the column-vector equation
\begin{equation}J(i\kappa)\zeta+\dot J(i\kappa)\alpha=0,\label{8.29}\end{equation}
where $\alpha\in \text{Ker}[J(i\kappa)]$ and $\zeta$ is some column vector
in ${\bf C}^n.$.
Let $\beta\in \text{Ker}[J(i\kappa)^\dagger]$ be the unique column vector corresponding to $\alpha$
as stated in Theorem~8.3. Thus we would have $J(i\kappa)^\dagger\beta=0$ or
equivalently
\begin{equation}\beta^\dagger J(i\kappa)=0.\label{8.30}\end{equation}
Let us premultiply (8.29) by $\beta^\dagger$ and use (8.30) in order to obtain
\begin{equation}\beta^\dagger \dot J(i\kappa)\alpha=0.\label{8.31}\end{equation}
Using Theorem~8.3 in (8.31) we see that we must have $\alpha=0$ and hence
$N_{-p}=0$ for $p\ge 2.$ Thus, from (8.27) we conclude that
$J(k)^{-1}$ must have a simple pole at $k=i\kappa.$ \qed

Having established that the expansion (8.27) contains
only a simple pole as
\begin{equation}J(k)^{-1}=
{\displaystyle}\frac{N_{-1}}{k-i\kappa}+N_0
+(k-i\kappa)\,N_1+O((k-i\kappa)^2),
\qquad k\to i\kappa,\label{8.32}\end{equation}
we would like investigate the term $N_{-1}$ further. One of our goals is
to relate the multiplicity of the bound state at $k=i\kappa$
to the multiplicity of the zero of $\det J(k)$ at $k=i\kappa$ and
to show that those two multiplicities are equal to each other.
Recall that the multiplicity $m_\kappa$ of the bound state at $k=i\kappa$ is
defined as the number of linearly independent column vectors that are
square-integrable column-vector solutions to (1.1) at $k=i\kappa$ and that also
satisfy the boundary condition (1.6)-(1.8). From (8.10) we see
that $m_\kappa$ is equal to the dimension of the kernel of
$J(i\kappa).$ Our goal is to prove that
$m_\kappa$ is also equal to the multiplicity of
the zero of $\det J(k)$ at $k=i\kappa.$

{\bf Theorem 8.5:} {\it Consider the selfadjoint matrix
Schr\"odinger operator with the selfadjoint boundary condition
(1.6)-(1.8) and with the potential $V$ satisfying (1.3) and (1.5).
Let $J(k)$
be the corresponding Jost matrix appearing in (3.1).
Assume that there is a bound state at $k=i\kappa$ for some positive $\kappa.$ Then, we have
\begin{equation}\det J(k)=c_3(k-i\kappa)^{m_\kappa}[1+O(k-i\kappa)],\qquad
k\to i\kappa,\label{8.33}\end{equation}
where $c_3$ is a nonzero constant and
$m_\kappa$ is the positive integer
appearing in (8.10) and denoting the multiplicity of the bound state at $k=i\kappa.$
Consequently, the order of the zero of $\det J(k)$ at $k=i\kappa$ is
equal to $m_\kappa.$}

{\it{Proof:}} From (8.10) we know that the geometric multiplicity
of the zero-eigenvalue of $J(i\kappa)$ is equal to $m_\kappa.$
We will proceed as in Section~6 of Ref.~\citenum{2}, and hence
we will omit some of the details by referring the reader
to Ref.~\citenum{2}. Using a similarity transformation
$J(i\kappa)\mapsto {\mathcal{S}}_1^{-1} J(i\kappa){\mathcal{S}}_1$
with an appropriate invertible matrix ${\mathcal{S}}_1,$
we will transform $J(i\kappa)$ to a Jordan canonical form.
Let us assume that there are
$\nu_\kappa$ Jordan chains and hence the Jordan canonical form of $J(i\kappa)$
contains $\nu_\kappa$ Jordan blocks. Let us use
$\lambda_s$ to denote the eigenvalue of
$J(i\kappa)$ associated with the $s$th Jordan chain,
where we realize that the eigenvalues may be repeated
and hence there may be more than one Jordan block
for a given eigenvalue $\lambda_s.$
Let us use $J_{n_s}(\lambda_s)$ to denote
the $s$th Jordan block, where we assume that the matrix
size of that block is $n_s\times n_s.$
Without loss of
generality we can assume that the first $m_\kappa$ Jordan chains
all belong to the zero eigenvalue of $J(i\kappa).$ Let us use
$\mu_\kappa$ to denote the algebraic multiplicity of
the zero eigenvalue of $J(i\kappa).$ Thus, we assume that the number of
nonzero eigenvalues (including multiplicities) of $J(i\kappa)$ is $n-\mu_\kappa.$
As a result, the first $m_\kappa$ Jordan blocks each have the form
\begin{equation}J_{n_s}(\lambda_s)=\left[ \begin{array}{cccccc}
0&1&0&\dots &0&0\\
0&0&1&\dots&0&0\\
\vdots&\vdots&\vdots&\ddots&\vdots&\vdots\\
0&0&0&\dots&0&1\\
0&0&0&\dots&0&0\end{array}\right],\qquad s=1,\dots,m_\kappa,\label{8.34}\end{equation}
and the remaining Jordan blocks associated with the nonzero eigenvalues
of $J(i\kappa)$
have the form
$$J_{n_s}(\lambda_s)=\left[ \begin{array}{cccccc}
\lambda_s&1&0&\dots &0&0\\
0&\lambda_s&1&\dots&0&0\\
\vdots&\vdots&\vdots&\ddots&\vdots&\vdots\\
0&0&0&\dots&\lambda_s&1\\
0&0&0&\dots&0&\lambda_s\end{array}\right],\qquad s=m_\kappa+1,\dots,\nu_\kappa,$$
with nonzero diagonal entries $\lambda_s.$
The Jordan canonical form of $J(i\kappa)$ is then given by
$${\mathcal{S}}_1^{-1}J(i\kappa)\,{\mathcal{S}}_1=\displaystyle\oplus_{s=1}^{\nu_\kappa} J_{n_s}(\lambda_s).$$
Next, let us move all the entries with $1$ appearing in the superdiagonal
in the first $m_\kappa$ Jordan blocks in (8.34)
and collect those entries into
the $(\mu_\kappa-m_\kappa)\times (\mu_\kappa-m_\kappa)$ identity
matrix $I_{\mu_\kappa-m_\kappa}.$ This can be achieved by using
the matrices $P_4$ and $P_5$ given by
$$P_4=\left[ \begin{array}{cc} \Pi_4&0\\
0&I_{n-\mu_\kappa}\end{array}\right],\quad
P_5=\left[ \begin{array}{cc} \Pi_5&0\\
0&I_{n-\mu_\kappa}\end{array}\right],$$
for some permutation matrices $\Pi_4$ and $\Pi_5$
that affect only the first $\mu_\kappa$ columns and
$\mu_\kappa$ rows, respectively, of the matrices on which they operate.
The combined matrix transformation
$J(i\kappa)\mapsto P_5{\mathcal{S}}_1^{-1} J(i\kappa){\mathcal{S}}_1 P_4$ results in
the upper-triangular matrix given by
\begin{equation}P_5{\mathcal{S}}_1^{-1} J(i\kappa){\mathcal{S}}_1 P_4=\text{diag}
\{0_{m_\kappa},I_{\mu_\kappa-m_\kappa},
J_{n_{\mu_\kappa+1}}(\lambda_{\mu_\kappa+1}),\dots
,J_{n_{\nu_\kappa}}(\lambda_{\nu_\kappa})\},\label{8.35}\end{equation}
where we recall that $0_{m_\kappa}$ denotes the $m_\kappa\times m_\kappa$ zero matrix.
Let us define the $(n-\mu_\kappa)\times (n-\mu_\kappa)$ matrix ${\bf d}_0$ as
\begin{equation}{\bf d}_0:=\text{diag}
\{I_{\mu_\kappa-m_\kappa},
J_{n_{\mu_\kappa+1}}(\lambda_{\mu_\kappa+1}),\dots
,J_{n_{\nu_\kappa}}(\lambda_{\nu_\kappa})\}.\label{8.36}\end{equation}
The matrix ${\bf d}_0$ is invertible
because it is an upper-triangular matrix with nonzero
diagonal entries.
Using (8.36) in (8.35) we obtain the block decomposition
\begin{equation}P_5{\mathcal{S}}_1^{-1} J(i\kappa){\mathcal{S}}_1 P_4=
\text{diag}
\{0_{m_\kappa},{\bf d}_0\}.\label{8.37}\end{equation}
Comparing (8.26) and (8.37) we see that
\begin{equation}P_5{\mathcal{S}}_1^{-1} J(k){\mathcal{S}}_1 P_4=
\text{diag}
\{0_{m_\kappa},{\bf d}_0\}+
(k-i\kappa)\left[ \begin{array}{cc} {\bf a}_1&{\bf b}_1\\
{\bf c}_1&{\bf d}_1 \end{array}\right]
+O((k-i\kappa)^2),\qquad k\to i\kappa,\label{8.38}\end{equation}
where we have let
$$\left[ \begin{array}{cc} {\bf a}_1&{\bf b}_1\\
{\bf c}_1&{\bf d}_1 \end{array}\right]:=P_5{\mathcal{S}}_1^{-1} \dot J(i\kappa){\mathcal{S}}_1 P_4.
$$
 From Theorem~8.4 we know that $J(k)^{-1}$ has a simple pole at $k=i\kappa$ and hence
with the help of (8.32) we get
\begin{equation}\left(P_5{\mathcal{S}}_1^{-1} J(k){\mathcal{S}}_1 P_4\right)^{-1}=
{\displaystyle}\frac{1}{k-i\kappa}\left[ \begin{array}{cc}
{\bf n}_1&{\bf n}_2\\
{\bf n}_3&{\bf n}_4\end{array}\right]+
\left[ \begin{array}{cc}
{\bf m}_1&{\bf m}_2\\
{\bf m}_3&{\bf m}_4\end{array}\right]+O(k-i\kappa),\qquad k\to i\kappa,\label{8.39}\end{equation}
where we have defined
$$\left[ \begin{array}{cc}
{\bf n}_1&{\bf n}_2\\
{\bf n}_3&{\bf n}_4\end{array}\right]:=P_4^{-1}{\mathcal{S}}_1^{-1} N_{-1} {\mathcal{S}}_1 P_5^{-1},\quad
\left[ \begin{array}{cc}
{\bf m}_1&{\bf m}_2\\
{\bf m}_3&{\bf m}_4\end{array}\right]:=P_4^{-1}{\mathcal{S}}_1^{-1} N_0 {\mathcal{S}}_1 P_5^{-1},\quad
$$
with $N_{-1}$ and $N_0$ being the matrices appearing in (8.32),
with some $m_\kappa\times m_\kappa$ block matrices ${\bf n}_1$ and ${\bf m}_1,$
some $(n-m_\kappa)\times (n-m_\kappa)$ block matrices
${\bf n}_4$ and ${\bf m}_4,$ and
the remaining block matrices of appropriate sizes.
Using (8.38) and (8.39) in the matrix identities
$$\begin{cases} \left(P_5{\mathcal{S}}_1^{-1} J(k){\mathcal{S}}_1 P_4\right)^{-1}
\left(P_5{\mathcal{S}}_1^{-1} J(k){\mathcal{S}}_1 P_4\right)=I_n,\\
$$\left(P_5{\mathcal{S}}_1^{-1} J(k){\mathcal{S}}_1 P_4\right)
\left(P_5{\mathcal{S}}_1^{-1} J(k){\mathcal{S}}_1 P_4\right)^{-1}=I_n,
\end{cases}$$
we obtain
\begin{equation}\left[ \begin{array}{cc}
{\bf n}_1&{\bf n}_2\\
{\bf n}_3&{\bf n}_4\end{array}\right]
\text{diag}
\{0_{m_\kappa},{\bf d}_0\}=0_n,
\quad
\text{diag}
\{0_{m_\kappa},{\bf d}_0\}
\left[ \begin{array}{cc}
{\bf n}_1&{\bf n}_2\\
{\bf n}_3&{\bf n}_4\end{array}\right]
=0_n,\label{8.40}\end{equation}
\begin{equation}\left[ \begin{array}{cc}
{\bf n}_1&{\bf n}_2\\
{\bf n}_3&{\bf n}_4\end{array}\right]
\left[ \begin{array}{cc}
{\bf a}_1&{\bf b}_1\\
{\bf c}_1&{\bf d}_1\end{array}\right]
+
\left[ \begin{array}{cc}
{\bf m}_1&{\bf m}_2\\
{\bf m}_3&{\bf m}_4\end{array}\right]
\text{diag}
\{0_{m_\kappa},{\bf d}_0\}=I_n,\label{8.41
}\end{equation}
\begin{equation}
\left[ \begin{array}{cc}
{\bf a}_1&{\bf b}_1\\
{\bf c}_1&{\bf d}_1\end{array}\right]
\left[ \begin{array}{cc}
{\bf n}_1&{\bf n}_2\\
{\bf n}_3&{\bf n}_4\end{array}\right]
+
\text{diag}
\{0_{m_\kappa},{\bf d}_0\}
\left[ \begin{array}{cc}
{\bf m}_1&{\bf m}_2\\
{\bf m}_3&{\bf m}_4\end{array}\right]
=I_n.\label{8.42}\end{equation}
Because ${\bf d}_0$ is invertible, from (8.40) we see that
\begin{equation}{\bf n}_2=0,\quad {\bf n}_3=0,\quad {\bf n}_4=0,\label{8.43}\end{equation}
for some zero matrices of appropriate sizes.
Using (8.43) in (8.41) and (8.42) we get
$${\bf n}_1 {\bf a}_1=I_{m_\kappa},
\quad {\bf n}_1 {\bf b}_1+{\bf m}_2 {\bf d}_0=0,
\quad {\bf m}_4 {\bf d}_0=I_{n-m_\kappa},\quad
{\bf c}_1 {\bf n}_1+ {\bf d}_0{\bf m}_3=0,$$
which establishes the invertibility of the block matrix
${\bf a}_1$ and also implies
\begin{equation}{\bf n}_1={\bf a}_1^{-1},\quad {\bf m}_2=-{\bf a}_1^{-1}{\bf b}_1 {\bf d}_0^{-1},
\quad {\bf m}_4={\bf d}_0^{-1},
\quad {\bf m}_3=-{\bf d}_0^{-1}{\bf c}_1{\bf a}_1^{-1}.\label{8.44}\end{equation}
Using (8.44) in (8.39) we
obtain the expansion
\begin{equation}
\left( P_5{\mathcal{S}}_1^{-1} J(k){\mathcal{S}}_1 P_4 \right)^{-1}=
\left[ \begin{array}{cc} \displaystyle\frac{{\bf a}_1^{-1}[I_{m_\kappa}+O(k-i\kappa)]
}{k-i\kappa}&-{\bf a}_1^{-1}
{\bf b}_1 {\bf d}_0^{-1} \\
-{\bf d}_0^{-1}{\bf c}_1  {\bf a}_1^{-1}
&{\bf d}_0^{-1}
\end{array}  \right]+O(k-i\kappa),
\qquad k\to i\kappa.\label{8.45}\end{equation}
 From (8.45) we see that
$$\det\left(\left(P_5{\mathcal{S}}_1^{-1} J(k){\mathcal{S}}_1 P_4\right)^{-1}
\right)={\displaystyle}\frac{\left(\det {\bf a}_1^{-1}\right) \left(\det {\bf d}_0^{-1}\right)}{(k-i\kappa)^{m_\kappa}}\left[1+O(k-i\kappa)\right],
\qquad k\to i\kappa,$$
or equivalently
\begin{equation}\det J(k)={\displaystyle}\frac{\det({\bf a}_1 {\bf d}_0)}{\det(P_4 P_5)}(k-i\kappa)^{m_\kappa}\left[1+O(k-i\kappa)\right],
\qquad k\to i\kappa.\label{8.46}\end{equation}
Thus, (8.46) establishes (8.33) with $c_3$ given by
$$c_3:=(\det P_4)(\det P_5)(\det{\bf a}_1)(\det {\bf d}_0),$$
which is nonzero due to the fact that $P_4$ and $P_5$ are some $n\times n$ permutation matrices and hence their
determinants are either $1$ or $-1,$
and the matrices ${\bf a}_1$ and ${\bf d}_0$ are invertible and hence
their determinants are nonzero. \qed

We remark the similarity between Theorem~8.5 and Corollary~6.2
and the similarity between (8.33) and (6.5).

Let us note that the transformation
specified in Proposition~4.1(a) on the boundary parameters
$A$ and $B$ does not affect the boundary condition (1.6)-(1.8).
This is because $(A,B)\mapsto (AT,BT)$
for an invertible matrix $T$ results in a premultiplication
of both sides of (1.6) by $T^\dagger$ as well as a premultiplication
by $T^\dagger$ and a postmultiplication by $T$ of both sides of
(1.7) and (1.8). Thus, as seen from (3.2), the potential $V$ and
the boundary parameters $A$ and $B$ cannot uniquely determine
the Jost matrix $J(k),$ but they determine
$J(k)$ uniquely up to a postmultiplication by an invertible matrix $T.$
However, such a nonuniqueness
does not affect the zeros in ${{\bf C}^+}$ of the determinant of
$J(k)$ because $\det J(k)$ and $\det[J(k)\,T]$ have the same set of zeros.
Hence, the bound states are not affected
by such a nonuniqueness, and the bound states
are uniquely determined by
the potential $V$ and the boundary parameters $A$ and $B$
appearing in (1.6)-(1.8).

The following result is relevant in establishing the finiteness
of the number of bound states.

{\bf Theorem 8.6:} {\it Consider
the Schr\"odinger operator corresponding to
(1.1) with the selfadjoint boundary condition
(1.6)-(1.8) and with the potential
$V$ satisfying (1.3) and (1.5). Let $J(k)$ be the corresponding
Jost matrix defined in (3.1). Then, the zeros of
$\text{det}\, J(k)$ in ${\overline{{\bf C}^+}}\setminus\{0\}$ can only occur
on the positive imaginary axis,
and
the number of such zeros, which we denote by $N$ (without counting multiplicities), is finite.}

{\it{Proof:}} By Theorem~3.1(a) we know that $J(k)$
is analytic in ${{\bf C}^+}$ and continuous in ${\overline{{\bf C}^+}}.$ Thus,
$\det J(k)$ possesses the same properties. Because of the selfadjointness
of the Schr\"odinger operator, the bound-state $k$-values, i.e.
the zeros of $\text{det} J(k)$ in ${\overline{{\bf C}^+}}\setminus\{0\},$
can occur either on the real axis or on the positive imaginary axis.
By Proposition~3.1(c), $J(k)$ is invertible for $k\in{\bf R}\setminus\{0\}$
and hence those zeros can only occur on the positive imaginary axis.
Let us use ${\mathcal{H}}$
to denote the set of zeros of $\det J(k)$ on the positive imaginary axis.
Because of (7.17), ${\mathcal{H}}$ is a bounded set.
Furthermore, (6.5) implies that $\det J(i\zeta)\ne 0$ for $0<\zeta<\kappa_0$
for some positive $\kappa_0$-value. Thus, it follows
that ${\mathcal{H}}\subset [i\kappa_0,ib]$ for some positive $b.$
We must prove that ${\mathcal{H}}$ is a finite set. If it were not a finite set,
being bounded, ${\mathcal{H}}$ would have to have an accumulation point
in $[i\kappa_0,ib].$ However, the analyticity of $\det J(k)$ in ${{\bf C}^+}$
would then require $\det J(k)\equiv 0$ in ${{\bf C}^+},$ contradicting (7.17). \qed

Let us assume that the $N$ distinct zeros
of $\text{det} J(k)$
on the positive imaginary axis occur at
$k=i\kappa_j$ with $j=1,\dots,N.$ If there are no bound states, then
we have $N=0.$ If there any bound states, as stated in Theorem~8.6, the positive integer
$N$ is finite.
Let us use $m_{\kappa_j}$ to denote the multiplicity
of the bound state at $k=i\kappa_j.$ As in (8.10), we have
$$m_{\kappa_j}=\text{dim}\,\text{Ker}[J(i\kappa_j)],$$
and hence $m_{\kappa_j}$ is a positive integer not exceeding $n.$
 From Theorems 8.1 and 8.6 we conclude that, the number of
bound states including the multiplicities, ${\mathcal{N}},$ is a finite number and
given by
\begin{equation}{\mathcal{N}}:=\sum_{j=1}^N m_{\kappa_j}.\label{8.47}\end{equation}
 From Theorem~8.5 it follows that
the multiplicity of the zero of $\det J(k)$ at $k=i\kappa_j$ is the same
as the multiplicity $m_{\kappa_j}$ of the bound state at $k=i\kappa_j.$ Thus,
 from (8.47) we have the following result.

{\bf Corollary 8.7:} {\it Consider
the Schr\"odinger operator corresponding to
(1.1) with the selfadjoint boundary condition
(1.6)-(1.8) and with the potential
$V$ satisfying (1.3) and (1.5). Let $J(k)$
be the corresponding Jost function given in (3.1), and let
${\mathcal{N}}$ be the corresponding number of
bound states (including multiplicities), as indicated in (8.47). Then, ${\mathcal{N}}$ is also equal to the
number of zeros (including multiplicities) of $\det J(k)$
in ${{\bf C}^+}.$}

In Section~9 we will relate ${\mathcal{N}}$ to the change in the argument
of the determinant of the scattering matrix $S(k)$ along the positive real axis.

\section{Levinson's theorem}

In this section we establish Levinson's theorem for the selfadjoint matrix
Schr\"odinger operator with the selfadjoint boundary condition
(1.6)-(1.8) and with the potential $V$ satisfying (1.3) and (1.5).
We do this by relating the argument of the determinant of the
scattering matrix $S(k)$ defined in
(3.8) to the number (including multiplicities) of bound states ${\mathcal{N}}$ given in (8.47).
We achieve our goal by applying the argument principle to
the determinant of the Jost function $J(k)$ given in (3.1).

Let us define $h(k)$ as
\begin{equation}h(k):=\det J(k).\label{9.1}\end{equation}
The region we will use in the argument principle
is the region whose boundary is
${\mathcal{C}}_{\epsilon,R},$ which consists of four pieces as given by
\begin{equation}{\mathcal{C}}_{\epsilon,R}:=(-R,-\epsilon)\cup{\mathcal{C}}_\epsilon\cup(\epsilon,R)\cup{\mathcal{C}}_R.\label{9.2}\end{equation}
The first piece $(-R,-\epsilon)$ is the directed line segment on the real axis
for some small positive $\epsilon$ and for a large positive $R,$
with the direction of the path from $-R+i0$ to $-\epsilon+i0.$
The second piece ${\mathcal{C}}_\epsilon$ consists of the
upper semicircle centered at the origin with radius
$\epsilon$ and traversed from the point $-\epsilon+i0$ to
the point $\epsilon+i0.$ The third piece $(\epsilon,R)$
is the directed line segment of the positive real axis from $\epsilon+i0$ to $R+i0.$
The fourth piece ${\mathcal{C}}_R$ is the
upper semicircle centered at the origin with radius
$R$ and traversed from the point $R+i0$ to
the point $-R+i0.$
The analyticity of $h(k)$ in our region and
its continuity in the closure of our region follows from Theorem~3.1(a).
By choosing $R$ large enough and by choosing $\epsilon$ small enough,
 from Theorem~8.6 we know that the only zeros
of $h(k)$ in our region can occur on the positive imaginary axis at $N$ distinct
points $k=i\kappa_j$
for some nonnegative integer $N$ and that $h(k)$ does
not vanish on the boundary of our region.

Let us use $\text{arg}[
h(k)]\big|_{{\mathcal{C}}}$ for the change in the argument of $h(k)$ along a path ${\mathcal{C}},$
and let us recall that an overdot indicates the $k$-derivative.

{\bf Proposition 9.1:} {\it Consider
the Schr\"odinger operator corresponding to
(1.1) with the selfadjoint boundary condition
(1.6)-(1.8) and with the potential
$V$ satisfying (1.3) and (1.5).
Let $J(k)$ and $S(k)$ be the corresponding Jost and scattering
matrices defined in (3.1) and (3.8), respectively.
Then, the change in
the argument of $\det S(k)$ along the directed path $(\epsilon,R)$
and the change in
the argument of $\det J(k)$ along the directed paths
$(-R,-\epsilon)$ and $(\epsilon,R)$
are related to each other as}
\begin{equation}\text{arg}[\det S(k)]\big|_{(\epsilon,R)}=-
\text{arg}[\det J(k)]\big|_{(\epsilon,R)}-\text{arg}[\det J(k)]\big|_{(-R,-\epsilon)}.\label{9.3}\end{equation}

{\it{Proof:}} From (3.8) we have
\begin{equation}\det S(k)=(-1)^n\,{\displaystyle}\frac{\det J(-k)}{
\det J(k)},\qquad k\in{\bf R}\setminus\{0\}.\label{9.4}\end{equation}
 From (3.11) we get
$|\det S(k)|=1$ and hence
\begin{equation}\det S(k)=e^{i\,\text{arg}[\det S(k)]},
\qquad k\in(\epsilon,R),\label{9.5}\end{equation}
\begin{equation}|\det J(-k)|=|\det J(k)|,\qquad k\in{\bf R}\setminus\{0\},\label{9.6}\end{equation}
where we have also used (9.4) to obtain (9.6).
By Theorem~3.1(c) we have $|\det J(k)|\ne 0$ for $k\in{\bf R}\setminus\{0\},$ and hence
with the help of (9.6) we get
\begin{equation}\det J(k)=|\det J(k)|\, e^{i\,\text{arg}[\det J(k)]},
\qquad k\in(\epsilon,R),\label{9.7}\end{equation}
\begin{equation}\det J(-k)=|\det J(k)|\, e^{i\,\text{arg}[\det J(-k)]},
\qquad k\in(\epsilon,R).\label{9.8}\end{equation}
Using (9.5), (9.7), and (9.8) in (9.4) we get
$$e^{i\,\text{arg}[\det S(k)]}=(-1)^{n}
e^{i\,\text{arg}[\det J(-k)]}e^{-i\,\text{arg}[\det J(k)]},
\qquad k\in(\epsilon,R),$$
 from which we obtain (9.3). \qed

In the following proposition, we provide the change in the argument of
$h(k)$ along the pieces of paths appearing in (9.2).

{\bf Proposition 9.2:} {\it Consider
the Schr\"odinger operator corresponding to
(1.1) with the selfadjoint boundary condition
(1.6)-(1.8) and with the potential
$V$ satisfying (1.3) and (1.5). Then, the function $h(k)$ defined in (9.1)
satisfies}
\begin{equation}\lim_{\epsilon\to 0^+}\lim_{R\to +\infty}
\int_{{\mathcal{C}}_{\epsilon,R}} dk\,{\displaystyle}\frac{\dot h(k)}{h(k)}
=2\pi i{\mathcal{N}},\label{9.9}\end{equation}
\begin{equation}\lim_{R\to +\infty}
\int_{{\mathcal{C}}_R}dk\, {\displaystyle}\frac{\dot h(k)}{h(k)}
=\pi i(n_{\text M}+n_{\text N}),\label{9.10}\end{equation}
\begin{equation}\lim_{\epsilon\to 0^+}
\int_{{\mathcal{C}}_\epsilon}dk\, {\displaystyle}\frac{\dot h(k)}{h(k)}
=-\pi i\mu,\label{9.11}\end{equation}
\begin{equation}
\int_{(-R,-\epsilon)\cup(\epsilon,R)}dk\, {\displaystyle}\frac{\dot h(k)}{h(k)}
=i\left(
\text{arg}[
h(k)]\big|_{(\epsilon,R)}+
\text{arg}[
h(k)]\big|_{(-R,-\epsilon)}\right)
,\label{9.12}\end{equation}
{\it where ${\mathcal{N}}$ is the nonnegative integer appearing in (8.47),
the paths ${\mathcal{C}}_{\epsilon,R},$ ${\mathcal{C}}_\epsilon,$ and
${\mathcal{C}}_R$ are those in (9.2),
$n_{\text M}$ and $n_{\text N}$ are the nonnegative integers defined after
(4.2), and $\mu$ is the (algebraic and geometric) multiplicity of
the eigenvalue $+1$ of the zero-energy
scattering matrix $S(0),$ with
$S(k)$ being the scattering matrix defined in (3.8).}

{\it{Proof:}} Because of Corollary~8.7, the number
of
bound states ${\mathcal{N}}$ (including multiplicities) is equal to
the number of
zeros of $\det J(k)$ (including multiplicities)
in ${{\bf C}^+}.$ We get (9.9) by applying the argument principle to $h(k)$ along
the closed path ${\mathcal{C}}_{\epsilon,R}$ and by using
the fact that ${\mathcal{N}}$ is the number of
zeros (including multiplicities) of $h(k)$ inside ${\mathcal{C}}_{\epsilon,R}.$
We note that (9.10) directly follows from (7.17), and
(9.11) directly follows from (6.5). Finally, (9.12) is obtained with the help
of (9.6). \qed

We next state Levinson's theorem.

{\bf Theorem 9.3:} {\it Consider
the Schr\"odinger operator corresponding to
(1.1) with the selfadjoint boundary condition
(1.6)-(1.8) and with the potential
$V$ satisfying (1.3) and (1.5). The number ${\mathcal{N}}$
of bound states (including multiplicities)
appearing in (8.47) is related to
the argument of the determinant of
the scattering matrix $S(k)$ defined in (3.8) as}
\begin{equation}\text{arg}[\det S(0^+)]-\text{arg}[\det S(+\infty)]=
\pi\left(2{\mathcal{N}}+\mu-n_{\text M}-n_{\text N}
\right),\label{9.13}\end{equation}
{\it where $\mu$ is the (algebraic and geometric) multiplicity of
the eigenvalue $+1$ of the zero-energy
scattering matrix $S(0),$
and
$n_{\text M}$ and $n_{\text N}$ are the nonnegative integers defined after
(4.2).}

{\it{Proof:}} By Proposition~3.3,
the determinant of $S(k)$
is continuous on ${\bf R}$ and hence the left-hand side in
(9.3) is given by
\begin{equation}\lim_{\epsilon\to 0^+}\lim_{R\to +\infty}
\left(\text{arg}[\det S(k)]\big|_{(\epsilon,R)}\right)
=-\text{arg}[\det S(0^+)]+\text{arg}[\det S(+\infty)],
\label{9.14}\end{equation}
By combining the results in (9.10)-(9.12), and
by using the relationships among (9.3), (9.12), and (9.14), we
evaluate the sum of the integrals in (9.10)-(9.12)
in the limit as $\epsilon\to 0^+$ and $R\to+\infty.$
That sum then must be equal to the value of the integral
given in (9.9),
resulting in (9.13). \qed

Let us comment on (9.13) and how it is related to
Levinson's theorem appearing in the literature elsewhere.
In our analysis of the selfadjoint Schr\"odinger
operator with the general selfadjoint boundary condition
(1.6)-(1.8), the unperturbed Hamiltonian is chosen to
satisfy the Neumann boundary condition. Such a choice is
compatible with the time-dependent derivation of the
scattering matrix and is motivated \cite{16} by applications in
quantum wires. One consequence of that choice
is apparent in the large-$k$ limits of $S(k).$ As seen from (5.7), (5.10) and
(7.20), we have
$S(k)=S(\infty)+O(1/k)$ as $k\to\pm\infty,$ with
\begin{equation}S(\infty)=M\,\text{diag}\{I_{n_{\text M}},-I_{n_{\text D}},I_{n_{\text N}}\}\,M^\dagger,\label{9.15}\end{equation}
where $M$ is the unitary matrix
appearing in (4.11).
As a result of (9.15), the argument of the determinant of
$S(k)$ as $k\to+\infty$ is given by
$$\lim_{k\to+\infty}\arg[\det S(k)]=\arg[\det S(\infty)]=(-1)^{n_{\text D}}\pi+2\pi j,\qquad
j=0,\pm 1,\pm 2,\dots.$$
One can choose a branch with $\arg[\det S(\infty)]=0$ if $n_{\text D}$ is even and
a branch with $\arg[\det S(\infty)]=\pi$ if $n_{\text D}$ is odd.
So, in the purely Dirichlet case, i.e. when
$n_{\text M}=n_{\text N}=0$ and $n_{\text D}=n,$ from (9.15) we get
$S(k)=-I_n+O(1/k)$ as $k\to\pm\infty.$
On the other hand, in the literature \cite{1,24,25} dealing solely
with the Dirichlet case, it
is customary to choose the unperturbed Hamiltonian
to satisfy the Dirichlet boundary condition, yielding
$S(k)=I_n+O(1/k)$ as $k\to\pm\infty.$
Hence, in the literature dealing solely
with the Dirichlet case, it is also customary
to use the particular branch of the argument
function  for
$\det S(k)$ in such a way that
that argument takes the zero value at $k=+\infty.$
In fact, in such a case, it is customary to let
$$\det S(k)=e^{2i \delta_S(k)},\qquad k\in(0,+\infty),$$
with $\delta_S(+\infty)=0.$ Then, Levinson's theorem
in such a case is given by
\begin{equation}\delta_S(0^+)=
\pi\left({\mathcal{N}}+{\displaystyle}\frac{\mu}{2}\right).\label{9.16}\end{equation}
In particular, in the scalar case, we have (9.16) with
$\mu=1$ in the exceptional case and $\mu=0$ in the generic case. \cite{24}

\noindent {\bf Acknowledgments.} The research leading to this
article was supported in part by Consejo Nacional de Ciencia y
Tecnolog\'{\i}a (CONACYT) under project CB2008-99100-F
and by the Department of Defense
under grant number DOD-BC063989.


\end{document}